\documentclass[10pt, conference]{IEEEtran}
\IEEEoverridecommandlockouts
\usepackage{cite}
\usepackage{amsmath,amssymb,amsfonts,amsthm}
\usepackage{algpseudocode}
\usepackage{graphicx}
\usepackage{textcomp}
\usepackage{xcolor}
\usepackage{xspace}
\usepackage{tikz}
\usetikzlibrary{calc}
\usepackage{pgfplots}
\usetikzlibrary{automata, positioning,patterns}
\usepackage{xstring}
\usepackage{multirow}
\usepackage[normalem]{ulem}
\usepackage[caption=false, font=footnotesize]{subfig}
\usepackage{diagbox}
\usepackage[super]{nth}
\usepackage[linesnumbered,ruled]{algorithm2e}
\usepackage{ulem}
\usepackage{siunitx}

\newcommand\eat[1]{}

\ifdefined\FINAL

\fi

\eat{typo list:
SDR-WiFi}

\eat{
1. eat the SNR measurement simulation
2. add a table in case study of taskset
3. eat the throughput and rtt experiments in submit ver
4. eat related work in submit ver
}

\pgfplotsset{
    table/search path={figures},
    compat=1.15
}

\newtheorem{definition}{Definition}
\newtheorem{theorem}{Theorem}

\begin{document}

\title{\LARGE RT-WiFi on Software-Defined Radio: Design and Implementation}

\author{
\IEEEauthorblockN{Zelin Yun\IEEEauthorrefmark{2}, Peng Wu\IEEEauthorrefmark{2}, Shengli Zhou\IEEEauthorrefmark{2}, Aloysius K. Mok\IEEEauthorrefmark{3}, Mark Nixon\IEEEauthorrefmark{4}, Song Han\IEEEauthorrefmark{2}}
\IEEEauthorblockA{\IEEEauthorrefmark{2}University of Connecticut}  
\IEEEauthorrefmark{2}Email: \{zelin.yun, peng.wu, shengli.zhou, song.han\}@uconn.edu
\IEEEauthorblockA{\IEEEauthorrefmark{3}University of Texas at Austin}
\IEEEauthorrefmark{3}Email: mok@cs.utexas.edu
\IEEEauthorblockA{\IEEEauthorrefmark{4}Emerson Automation Solutions}  
\IEEEauthorrefmark{4}Email: mark.nixon@emerson.com
}

\maketitle
\thispagestyle{plain}
\pagestyle{plain}

\begin{abstract}
Applying high-speed real-time wireless technologies in industrial applications has the great potential to reduce the deployment and maintenance costs compared to their wired counterparts. Wireless technologies enhance the mobility and reduce the communication jitter and delay for mobile industrial equipment, such as mobile collaborative robots. Unfortunately, most existing wireless solutions employed in industrial fields either cannot support the desired high-speed communications or cannot guarantee deterministic, real-time performance. A more recent wireless technology, RT-WiFi, achieves a good balance between high-speed data rates and deterministic communication performance. It is however developed on commercial-of-the-shelf (COTS) hardware, and takes considerable effort and hardware expertise to maintain and upgrade. To address these problems, this paper introduces the software-defined radio (SDR)-based RT-WiFi solution which we call SRT-WiFi. SRT-WiFi provides full-stack configurability for high-speed real-time wireless communications. We present the overall system architecture of SRT-WiFi and discuss its key functions which achieve better timing performance and solve the queue management and rate adaptation issues compared to COTS hardware-based RT-WiFi. To achieve effective network management with rate adaptation in multi-cluster SRT-WiFi, a novel scheduling problem is formulated and an effective algorithm is proposed to solve the problem. A multi-cluster SRT-WiFi testbed is developed to validate the design, and extensive experiments are performed to evaluate the performance at both device and system levels.
\end{abstract}

\vspace{0.05in}

\section{Introduction}\label{sec:intro}

A recent trend in smart factory automation is to employ high-speed real-time wireless technologies to interconnect heterogeneous industrial assets to perform various sensing and control services, and support mobile equipment to conduct designated tasks in a collaborative fashion~\cite{tramarin2019real}. Most of these industrial applications have stringent requirements on both high data throughput and deterministic real-time performance ({\em e.g.}, latency and jitter)~\cite{guo2021towards,sisinni2018industrial}.

The existing efforts on the design and implementation of real-time wireless solutions can be summarized in four main categories. The first category includes those works focusing on low-speed low-power real-time communication solutions, such as WirelessHART, ISA 100.11a, WISA and 6TiSCH~\cite{song2008complete, isa10011a, steigmann2006introduction, dujovne20146tisch}. Although those solutions can achieve deterministic communication performance and have ultra-low energy footprint, they cannot support high-speed communications, constrained by the underlying IEEE 802.15.4/802.15.4e~\cite{802154wg} physical layer (PHY) and data link layer (DLL). The works in the second category~\cite{cheng2019adopting, cena2010performance, seno2012tuning, tian2016deadline, trsek2006simulation, moraes2007vtp} achieve the real-time performance based on IEEE 802.11e standard, including the hybrid coordination function (HCF) controlled channel access (HCCA) which enables the polling method~\cite{lam2015fast, lin2015polling, zheng2009industrial} and the enhanced distributed channel access (EDCA) which enables priorities in the transmissions and uses the highest priority for the real-time transmissions to guarantee their access to the channel. However, when EDCA is applied, the downlinks may compete for the highest priority queue on the access point (AP) side which may cause unnecessary delay and the ensuing timing violations. The {HCCA-based} polling method is not time-efficient when the channel usage is high compared to assigning the communication schedules to the devices directly and it is also subject to coexistence issues in the scenarios when multiple APs use the same HCCA access function~\cite{tramarin2019real}. The works in the third category study the applications of 5G and Long Term Evolution (LTE) technologies in real-time industrial applications~\cite{holfeld2016wireless, ashraf2016ultra, aijaz2020private}. However, the deployment of LTE and 5G equipment do not exploit the license-free bands and therefore misses the economic advantage and the flexibility afforded by the extra bandwidth required for the anticipated applications in the industrial automation field such as robotics. For the last category, existing works focus on modifying IEEE 802.11 standards and implementing the systems on commercial-of-the-shelf (COTS) hardware. For example, \cite{wei2013rt} proposes a configurable real-time WiFi system, called RT-WiFi, based on COTS hardware AR9285. It modifies the driver and implements a network manager for scheduling deterministic real-time communications. \cite{liang2019wia} proposes the WIA-FA system for wireless factory automation.  However, the uplinks of WIA-FA are {contention-based} and thus the transmissions from stations to APs are not deterministic. For the above works using COTS hardware, a major issue is that COTS hardware is usually not open-source, and many functions are not accessible which makes it difficult to maintain and upgrade such system to support frequently updated OS kernels and wireless protocols.

\begin{figure}[t]
\centering
    \includegraphics[width=1\columnwidth]{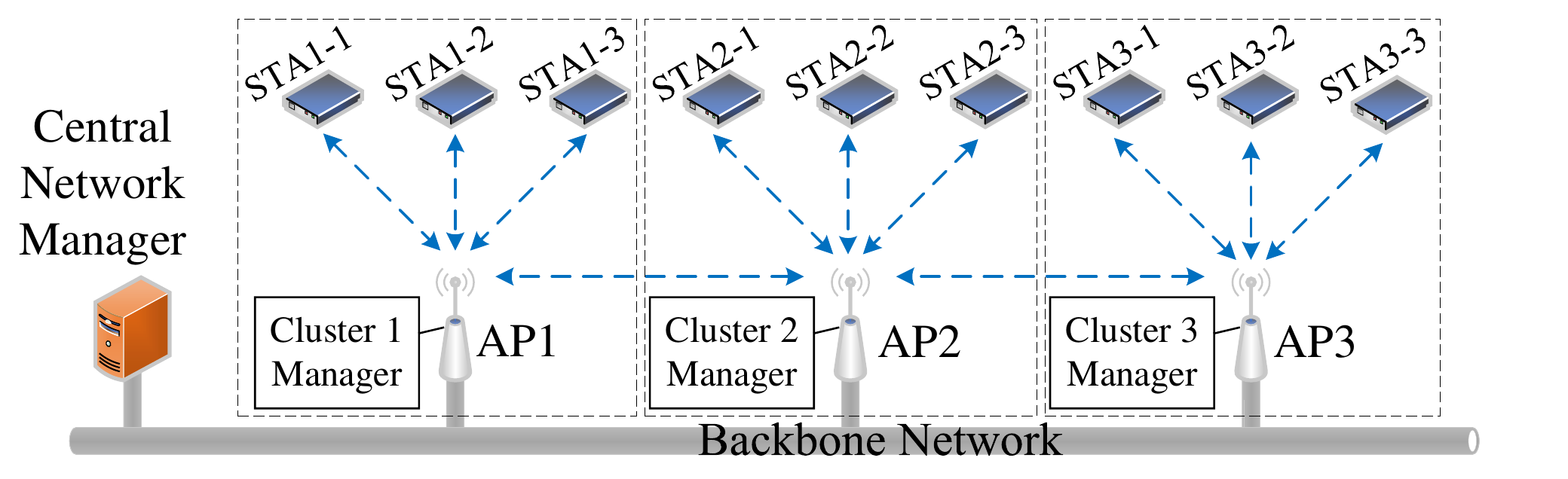}
    \vspace{-0.2in}
    \caption{Overview of the multi-cluster SRT-WiFi network.}
    \label{fig:systemmodel}
    \vspace{-0.1in}
\end{figure}

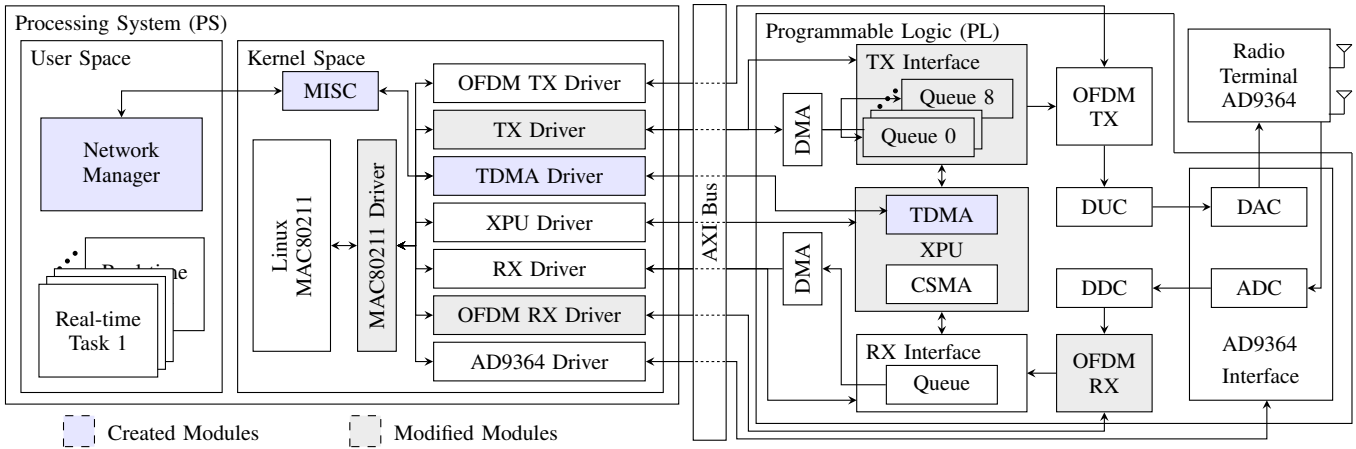
\begin{figure*}[t]
\centering
    \resizebox{1\hsize}{!}
    {\begin{tikzpicture}[every node/.style={font=\footnotesize},>=stealth]


\draw (9.9,5.5) node [anchor=north west] (pl){Programmable Logic (PL)} rectangle (17.6,0.2)[fill=white];
\path [fill=white] (17.7,3.7) rectangle (15.3,5.6);
\draw [-] (15.3,3.7)--(15.3,5.5);
\draw [-] (15.3,3.7)--(17.6,3.7);

\node[draw,text width=1.6cm, align=center, minimum height=1.2cm] (radioterm) at (16.4,4.7) {Radio\\Terminal\\AD9364};
\draw [-] (17.3,4.2)--(17.5,4.2);
\draw [-] (17.5,4.4)--(17.5,4.2);
\draw [-] (17.5,4.4)--(17.4,4.5);
\draw [-] (17.6,4.5)--(17.4,4.5);
\draw [-] (17.6,4.5)--(17.5,4.4);

\draw [-] (17.3,4.8)--(17.5,4.8);
\draw [-] (17.5,5)--(17.5,4.8);
\draw [-] (17.5,5)--(17.4,5.1);
\draw [-] (17.6,5.1)--(17.4,5.1);
\draw [-] (17.6,5.1)--(17.5,5);

\draw (15.5,3.5) node [anchor=north west] (adintf){} rectangle (17.35,0.5)[fill=white];
\draw (15.8,1) node [anchor=south west]{AD9364};
\draw (15.8,0.6) node [anchor=south west]{Interface};

\node[draw,text width=1.0cm, align=center, minimum height=0.5cm] (adc) at (16.4,1.95) {ADC};
\node[draw,text width=1.0cm, align=center, minimum height=0.5cm] (dac) at (16.4,3) {DAC};

\node[draw,text width=1.0cm, align=center, minimum height=0.5cm] (duc) at (14.4,3) {DUC};
\node[draw,text width=1.0cm, align=center, minimum height=0.5cm] (ddc) at (14.4,1.95) {DDC};

\node[draw,text width=1.0cm, align=center, minimum height=1.0cm] (ofdmtx) at (14.4,4.3) {OFDM\\TX};
\node[draw,text width=1.0cm, align=center, minimum height=1.0cm, fill=gray!15] (ofdmrx) at (14.4,0.85) {OFDM\\RX};

\node[draw,text width=2.0cm, align=center, minimum height=1.6cm, fill=gray!15] (xpu) at (12.3,2.45) {XPU};


\draw (11.2,5.1) node [anchor=north west] (txintf){TX Interface} rectangle (13.4,3.55)[fill=gray!15];
\node[draw,text width=1.2cm, align=center, minimum height=0.5cm][fill=gray!15] (txqueue8) at (12.5,4.4) {Queue 8};
\draw (11.5,4.3) circle (0.025)[fill=black];
\draw (11.7,4.5) circle (0.025)[fill=black];
\draw (11.6,4.4) circle (0.025)[fill=black];
\node[draw,text width=1.2cm, align=center, minimum height=0.5cm][fill=gray!15] (txqueue1) at (12.1,4) {Queue 1};
\node[draw,text width=1.2cm, align=center, minimum height=0.5cm][fill=gray!15] (txqueue0) at (12,3.9) {Queue 0};

\draw (11.2,1.35) node [anchor=north west] (rxintf){RX Interface} rectangle (13.4,0.35)[fill=white];
\node[draw,text width=1.2cm, align=center, minimum height=0.5cm][fill=white] (rxqueue) at (12.3,0.7) {Queue};

\node[draw,text width=1.2cm, align=center, minimum height=0.5cm, fill=blue!10] (tdma) at (12.3,2.9) {TDMA};
\node[draw,text width=1.2cm, align=center, minimum height=0.5cm, fill=white] (csma) at (12.3,2) {CSMA};

\node[draw,text width=0.7cm, align=center, minimum height=0.5cm, rotate=90] (dmatx) at (10.5,4) {DMA};
\node[draw,text width=0.7cm, align=center, minimum height=0.5cm, rotate=90] (dmarx) at (10.5,2.2) {DMA};

\draw (0.2,5.6) node [anchor=north west] (ps){Processing System (PS)} rectangle (8.9,0.45)[fill=white];

\draw (3.2,5.15) node [anchor=north west] (ks){Kernel Space} rectangle (8.7,0.6)[fill=white];

\node[draw,text width=2.5cm, align=center, minimum height=0.5cm] (otxd) at (7.1,4.6) {OFDM TX Driver};
\node[draw,text width=1.0cm, align=center, minimum height=0.5cm, fill=blue!10] (misc) at (4.4,4.5) {MISC};
\node[draw,text width=2.5cm, align=center, minimum height=0.5cm, fill=gray!15] (txd) at (7.1,4) {TX Driver};
\node[draw,text width=2.5cm, align=center, minimum height=0.5cm, fill=blue!10] (tdmad) at (7.1,3.4) {TDMA Driver};
\node[draw,text width=2.5cm, align=center, minimum height=0.5cm] (xpud) at (7.1,2.8) {XPU Driver};
\node[draw,text width=2.5cm, align=center, minimum height=0.5cm] (rxd) at (7.1,2.2) {RX Driver};
\node[draw,text width=2.5cm, align=center, minimum height=0.5cm, fill=gray!15] (orxd) at (7.1,1.6) {OFDM RX Driver};
\node[draw,text width=2.5cm, align=center, minimum height=0.5cm] (add) at (7.1,1) {AD9364 Driver};

\node[draw,text width=2.5cm, align=center, minimum height=0.5cm, rotate=90, fill=gray!15] (opend) at (5,2.5) {MAC80211 Driver};
\node[draw,text width=2.5cm, align=center, minimum height=1.0cm, rotate=90] (mac) at (3.9,2.5) {Linux\\MAC80211};

\draw (0.4,5.15) node [anchor=north west] (us){User Space} rectangle (3,0.6)[fill=white];
\node[draw,text width=1.85cm, align=center, minimum height=1.2cm,fill=blue!10] (nm) at (1.7,3.55) {Network\\Manager};
\node[draw,text width=1.3cm, align=center, minimum height=1.2cm][fill=white] (rttaskn) at (2,2) {Real-time\\Task n};
\draw (0.9,2.2) circle (0.025)[fill=black];
\draw (1.1,2.4) circle (0.025)[fill=black];
\draw (1,2.3) circle (0.025)[fill=black];
\node[draw,text width=1.3cm, align=center, minimum height=1.2cm][fill=white] (rttaskn3) at (1.6,1.6) {Real-time\\Task 3};
\node[draw,text width=1.3cm, align=center, minimum height=1.2cm][fill=white] (rttaskn2) at (1.5,1.5) {Real-time\\Task 2};
\node[draw,text width=1.3cm, align=center, minimum height=1.2cm][fill=white] (rttaskn1) at (1.4,1.4) {Real-time\\Task 1};

\draw[<-](ofdmtx)--+(-1,0);
\draw[->](ofdmtx)--(duc);
\draw[->](duc)--(dac);
\draw[<-](ddc)--(adc);
\draw[<-](ofdmrx)--(ddc);
\draw[->](ofdmrx)--+(-1,0);
\draw[<->](xpu)--+(0,1.1);
\draw[<->](xpu)--+(0,-1.1);

\draw[->](dmatx)--+(0.5,0)--+(0.5,-0.1)--(txqueue0);
\draw[-](dmatx)--+(0.5,0)--+(0.5,0)--+(0.8,0);
\draw[->](dmatx)--+(0.5,0)--+(0.5,0.4)--(txqueue8);

\draw[<-](dmarx)--+(0.5,0)--+(0.5,-1.5)--+(1.05,-1.5)--(rxqueue);

\draw[<->](rxd)--+(2.95,0)--+(2.95,-1.7)--+(4.1,-1.7);
\draw[<-](rxd)--(dmarx);

\draw[<->](ofdmrx)--+(0,-0.75)--+(-4.6,-0.75)--+(-4.6,0.75)--(orxd);

\draw[<->](add)--+(2.55,0)--+(2.55,-1)--+(9.4,-1)--+(9.4,-0.5);

\draw[<->](xpud)--+(2.9,0)--+(2.9,0)--+(4.1,0);

\draw[<->](tdmad)--+(3.05,0)--+(3.05,-0.45)--+(4.5,-0.45);

\draw[<->](txd)--+(2.7,0)--+(2.7,0.9)--+(4.1,0.9);
\draw[->](txd)--(dmatx);

\draw[<->](otxd)--+(2.55,0)--+(2.55,1)--+(7.3,1)--(ofdmtx);

\node[draw,text width=5.4cm, align=center, minimum height=0.4cm, rotate=90, fill=white] (axi) at (9.3,2.8) {AXI Bus};
\draw[dash pattern={on 1pt off 1pt on 1pt off 1pt}](9.5,1)--(9.1,1);
\draw[dash pattern={on 1pt off 1pt on 1pt off 1pt}](9.5,3.4)--(9.1,3.4);
\draw[dash pattern={on 1pt off 1pt on 1pt off 1pt}](9.5,2.2)--(9.1,2.2);
\draw[dash pattern={on 1pt off 1pt on 1pt off 1pt}](9.5,1.6)--(9.1,1.6);
\draw[dash pattern={on 1pt off 1pt on 1pt off 1pt}](9.5,2.8)--(9.1,2.8);
\draw[dash pattern={on 1pt off 1pt on 1pt off 1pt}](9.5,4)--(9.1,4);
\draw[dash pattern={on 1pt off 1pt on 1pt off 1pt}](9.5,4.6)--(9.1,4.6);

\draw[<->](opend)--+(0.5,0)--+(0.5,2.1)--(otxd);
\draw[<->](opend)--+(0.5,0)--+(0.5,1.5)--(txd);
\draw[<->](opend)--+(0.5,0)--+(0.5,0.9)--(tdmad);
\draw[<->](opend)--+(0.5,0)--+(0.5,0.3)--(xpud);
\draw[<->](opend)--+(0.5,0)--+(0.5,-0.3)--(rxd);
\draw[<->](opend)--+(0.5,0)--+(0.5,-0.9)--(orxd);
\draw[<->](opend)--+(0.5,0)--+(0.5,-1.5)--(add);

\draw[<->](mac)--(opend);

\draw[<->](nm)--+(0,0.95)--(misc);
\draw[<->](misc)--+(1,0)--+(1,-1.1)--(tdmad);

\draw[->](dac)--(radioterm);
\draw[<-](adc)--+(0.8,0)--+(0.8,2.15);

\draw[draw, dashed, fill=blue!10] (0.95,-0.1) rectangle +(0.4,0.4) node [pos=.5] {};
\draw (1.4,0.3) node [anchor=north west] {Created Modules};
\draw[draw, dashed, fill=gray!15] (4.65,-0.1) rectangle +(0.4,0.4) node [pos=.5] {};
\draw (5.1,0.3) node [anchor=north west] {Modified Modules};
\end{tikzpicture}}
    \vspace{-0.2in}
    \caption{Overview of the SRT-WiFi system architecture design based on the Openwifi project.}
    \label{fig:arch}
    \vspace{-0.1in}
\end{figure*}

To address the aforementioned issues in existing works, we present in this paper the design and implementation of a {software-defined radio (SDR)-based} RT-WiFi solution which we name SRT-WiFi. SDR~\cite{dillinger2005software} is a radio communication system where components that have been traditionally implemented in hardware are instead implemented by means of software on a PC or an embedded system. We design SRT-WiFi based on an advanced SDR platform (ZC706 development board with Zynq-7000 and AD9364) where the radio functions are programmed on field programmable gate array (FPGA). This advanced SDR system can run in real time since the radio functions are achieved by the logic blocks in FPGA running at the speed as driven by an oscillator. With such a programmable real-time radio system, SRT-WiFi can achieve the key functions required to support high-speed real-time communications, and also provide an open-source platform to support ever-evolving IEEE 802.11 standards.

Fig.~\ref{fig:systemmodel} shows the overview of a multi-cluster SRT-WiFi network where multiple APs are synchronized and connected to a backbone network. A central network manager (CNM) manages all the network resources and allocates them to the cluster managers (CMs) running on individual APs. In each cluster, high-speed real-time point-to-point wireless communications with rate adaptation are supported to deal with the interfered environments. The clusters operate on multiple channels meaning that one channel has one or multiple clusters operating on it. Compared to COTS hardware-based existing works, SRT-WiFi leverages the programmability of the SDR-based PHY and DLL to provide full-stack configurability.\footnote{The current version of SRT-WiFi system supports IEEE 802.11a/g. It can be further extended to support emerging IEEE 802.11 standards, such as 802.11n/ac/ax. See the ongoing and future work in Section~\ref{sec:conclusion}).} By taking advantage of this full-stack configurability, it is possible to add three major features in SRT-WiFi: i) more precise time synchronization which leads to a smaller slot size for packet transmission and higher sampling rate; ii) efficient queue management which reduces possible downlink latency caused by the limited number of queues in COTS hardware; iii) more accurate signal-to-noise ratio (SNR) measurement, based on which we propose a novel rate adaptation mechanism to dynamically change the data rates based on the SNR measurement of the links to guarantee the desired packet delivery ratio (PDR) of each link; this adaptation outperforms the Minstrel algorithm~\cite{xia2013evaluation} employed in regular WiFi network. Based on the proposed rate adaptation mechanism, we further formulate and solve the multi-cluster SRT-WiFi network scheduling problem (MSNS-RA) based on the dynamic rates determined at run time. We implement the SRT-WiFi protocol and the multi-cluster SRT-WiFi network management solution on a real testbed. Our extensive experiment results validate the effectiveness of the designs, and we evaluate the performance of SRT-WiFi at both device and system levels.

The remainder of this paper is organized as follows. Section~\ref{sec:arch} presents the overall system architecture of SRT-WiFi. Section~\ref{sec:pl} and Section~\ref{sec:ps} describe the design of the programmable logic (PL) component and the processing system (PS) component of SRT-WiFi, respectively. Section~\ref{sec:schedule} introduces the multi-cluster network management framework to support rate adaptation in SRT-WiFi to guarantee the timing requirement of real-time tasks even in the presence of severe interference. Section~\ref{sec:performance} evaluates the performance of SRT-WiFi at both device and system levels. Section~\ref{sec:relatedwork} gives a summary of the related work. We conclude the paper in Section~\ref{sec:conclusion} and discuss the ongoing and future work.

\section{System Architecture}\label{sec:arch}
We now present the overall architecture of SRT-WiFi system (see Fig.~\ref{fig:arch}). SRT-WiFi is based on the Openwifi project~\cite{Openwifigithub} which is a SoftMAC IEEE 802.11 design compatible with Linux MAC80211. We first introduce Openwifi, and then describe the SRT-WiFi architecture in detail.

\subsection{Openwifi Architecture}
Openwifi has two major components: the Processing System (PS) and the Programmable Logic (PL). PS is an operating system (OS) running the major part of the data link layer (DLL) and all the other higher layers. PL is an FPGA-based embedded system running the real-time part of the DLL and the physical layer (PHY). Both PL and PS are implemented on an integrated System-on-Chip (SoC) Zynq-7000~\cite{crockett2014zynq} which consists of an FPGA (for PL) and an ARM processor (for PS).  PL and PS exchange data through the Advanced eXtensible Interface (AXI)~\cite{math2011data} bus which supports direct memory access (DMA) and register reading and writing. In addition, PL connects to a radio terminal (AD9364~\cite{ad9364}) for packet transceiving.

In Openwifi, PL is designed as the wireless adaptor. As shown on the right side of Fig.~\ref{fig:arch}, PL has three main modules: the TX interface (TXI), the XPU (application-specific processing unit) and the RX interface (RXI). The TXI and RXI modules handle packet transmission and reception, respectively. The XPU module runs the state machine of IEEE 802.11 channel access methods, e.g., the distributed coordination function (DCF)~\cite{80211wg}. To process general packet transmission, TXI first holds the packet passed from PS (through direct memory access (DMA)) in its queues and waits for the transmission trigger from XPU. The carrier-sense multiple access (CSMA) block in XPU senses the channel and runs the backoff mechanism. Once the channel is available, XPU triggers TXI which in turn fires the packet to the modulation block (OFDM TX) and the digital up converter (DUC) module. The processed digital signal is then passed to the radio terminal through the digital-to-analog converter (DAC) block and is finally emitted from the antenna by the radio terminal. After sending the packet, the XPU module waits for the acknowledgement (ACK) packet from RXI if ACK is required. If ACK is correctly received or not required, XPU notifies TXI that the packet is transmitted successfully. TXI further triggers an interruption to PS to report the result. If XPU does not receive the correct ACK past a pre-defined time threshold, it triggers retransmission(s) until reaches the limit of transmission attempts and then reports the failure to PS. For a general packet reception, the terminal provides the received digital signal to PL through analog-to-digital (ADC) interface. After signal passing through the digital down converter (DDC) and demodulated by OFDM RX, RXI puts the packet in a queue. At the same time, XPU reads the packet header and applies a packet filter to decide if this packet is destined for PS. If so, RXI then fires the packet to PS. If the packet requires an ACK, XPU generates the ACK packet in TXI and triggers the transmission. Please note that both TX and RX of the radio terminal run in parallel. And all PL modules have registers to be used to configure the operation mode and parameters.

The PS component in Openwifi is a Linux OS running on an ARM processor. For being a SoftMAC wireless device in Linux, the major part of DLL is integrated in Linux kernel (MAC80211 subsystem~\cite{murlinux}) except for the real-time part of DLL and PHY which are implemented in PL. Thus the MAC80211 driver is needed to provide the interface of the wireless adaptor (PL) for the Linux MAC80211 subsystem. The data exchanges between the MAC80211 driver and PL rely on the sub-drivers (see the left side of Fig.~\ref{fig:arch}). All the sub-drivers are designed to provide APIs for register reading and writing to the MAC80211 driver so that it is able to configure the PL. The TX and RX drivers handle TX and RX data packet transfer between PS and PL, respectively, through DMA.

\subsection{SRT-WiFi Architecture}
The key design goal of SRT-WiFi is to support precise time synchronization and multi-cluster real-time communications with effective rate adaptation at run time. For this purpose, we present below the SRT-WiFi architecture, by modifying the PL and PS components in Openwifi to add the required functions.

\vspace{0.025in}
\noindent {\bf SRT-WiFi PL:}
The PL component of SRT-WiFi is designed to i) achieve the real-time transmissions with high synchronization time precision, ii) provide more efficient queue management and iii) measure the reception SNR of the links more precisely in order to provide reference for rate adaptation.

To achieve real-time transmissions, we design a TDMA block in XPU to supplement the CSMA block. The TDMA block triggers the PHY and DLL activities with high time precision. It runs either according to the local timer or synchronizes with another device in the SRT-WiFi network. According to our measurements (to be elaborated later), the synchronization time error and standard deviation in the multi-cluster SRT-WiFi are as low as 0.73 \SI{}{\micro\second} and 0.1 \SI{}{\micro\second}, respectively. Different from the CSMA block which triggers the transmissions following the DCF mechanism, the TDMA block triggers the transmission according to a schedule constructed by the network managers in PS. The schedule is stored in the TDMA block and updated at run time through a TDMA driver that is added in PS (see Fig.~\ref{fig:arch}). The TDMA and CSMA modes in SRT-WiFi can be switched during the run time seamlessly.

For an AP working in the TDMA mode, it needs to handle the links to all the connected stations. The transmissions on those links have to follow the order of a schedule. With limited number of queues, COTS hardware~\cite{wei2013rt} must manage the issue that the queued packets may block the transmissions of upcoming packets that may cause unmanageable congestion as the network scales up.  SRT-WiFi provides an effective queue management in the following two aspects: i) More queues are implemented in the TXI module and judiciously assigned to the real-time communication links to avoid delays, ii) A dynamic method is developed to allow the links to share the queues to further increase the number of supported links.

By leveraging the capability of SRT-WiFi to have direct access to the received signals, we are able to design novel methods to measure the SNR precisely and implement it in the OFDM RX module. The SNR information provides a reference for the rate adaptation mechanism to adapt the TX data rates and adjust the communication schedules at run time.

\vspace{0.025in}
\noindent {\bf SRT-WiFi PS Kernel:}
As shown in Fig.~\ref{fig:arch}, we add the TDMA driver and modify the MAC80211 driver, TX and OFDM RX drivers to provide an interface for exchanging the schedule, queue and SNR information between PS and PL in the TDMA mode. The TDMA driver is registered in the kernel as a miscellaneous character driver (MISC). It provides APIs for the network managers in the user space. The network manager configures the schedule and queue information in PL through the TDMA driver and updates the data rates in MAC80211 driver as well. The TX driver is modified to support queue management and the OFDM RX driver is enhanced to support reading the SNR values measured in PL.

\vspace{0.025in}
\noindent {\bf SRT-WiFi Network Management:}
We call the network managers running on individual APs cluster managers (CM) and the ones running on the stations device managers (DM). These network managers are designed for two purposes. The first one is to exchange information at the application layer among all the devices in the SRT-WiFi network. Those information includes the schedule, data rates and SNR of links in the SRT-WiFi network. The second purpose is to manage the TDMA DLL on each device such as configuring the schedule for the TDMA block and reading the SNR measurement from the PL. The device managers send SNR values to a central network manager (CNM) and CNM decides the schedule with a heuristic multi-cluster scheduling algorithm and returns the computed schedule and data rates according to the delay and throughput requirements of all the links so that we can support the dynamic rate adaptation to provide stable communications for multi-cluster SRT-WiFi. All network managers run in the user space so they are easy to maintain and upgrade.

\section{SRT-WiFi Programmable Logic (PL) Design}\label{sec:pl}

\begin{figure*}[t]
\centering
  \includegraphics[width=0.99\textwidth]{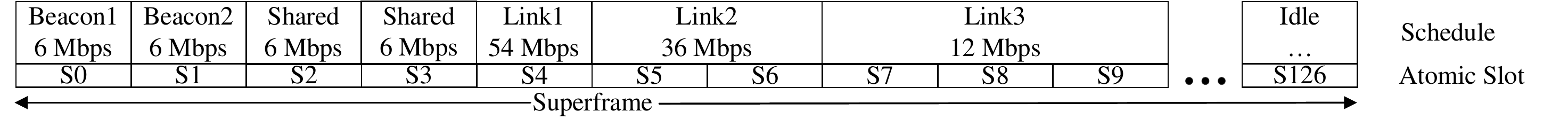}
\vspace{-0.2in}
\caption{The timing diagram of an example superframe in multi-cluster SRT-WiFi with 127 atomic slots.}
\label{fig:superframe}
\end{figure*}

We first introduce the PL design of SRT-WiFi, focusing on the XPU, TXI and OFDM RX modules. The TDMA-based data link layer (DLL) is designed as part of the XPU to achieve real-time performance. The queue management function is achieved in TXI and the SNR measurement function is implemented in OFDM RX module. In a multi-cluster SRT-WiFi network, the master AP (MAP) serves as the reference clock for the network. The slave APs (SAPs) synchronize with the MAP and the stations synchronize with their corresponding APs. The synchronization function of both APs and stations are implemented in their XPU modules. In the following, we elaborate on the design details of these functions.

\subsection{TDMA Block Design}
In the TDMA mode of SRT-WiFi, we aim to transceive frames at specified times to coordinate the communications between APs and stations to avoid collision. For this aim, all transmissions follow a schedule. The schedule describes the transmitting times and orders of the links in a time period called the superframe. The superframe is a sequence of consecutive time slots, and each time slot specifies the radio activities (TX, RX or Idle) and the associated sender/receiver. At run time, the superframe is repeated {\em ad infinitum} to schedule the transmissions. To support rate adaptation, the length of the time slot varies along with the rate, since with the same packet length, a lower rate requires longer time for transmission. The time slots use atomic slots as the basic time unit. In SRT-WiFi, the lengths of superframe, time slot and atomic slot are all configurable. The superframe length is mainly decided by the application requirements and the time/atomic slot lengths are decided by the selected data rates.

Fig.~\ref{fig:superframe} shows the timing diagram of an example superframe in an SRT-WiFi network. It has 127 atomic slots where Slot0 and Slot1 are used by AP1 and AP2 to send beacons, respectively. Slot2 and Slot3 are shared slots for any links and usually used for the association process. The other atomic slots are either assigned to links for dedicated communications or left idle. The links that use the same MTU but different data rates will require different slot lengths in terms of the number of atomic slots. For example, Link1, Link2 and Link3 use 1, 2 and 3 atomic slots for their transmissions, respectively.\footnote{The detailed data rates and their corresponding required slot lengths are presented in Section~\ref{sec:performance}.}

\begin{figure}[h]
    \centering
    \includegraphics[width=1\columnwidth]{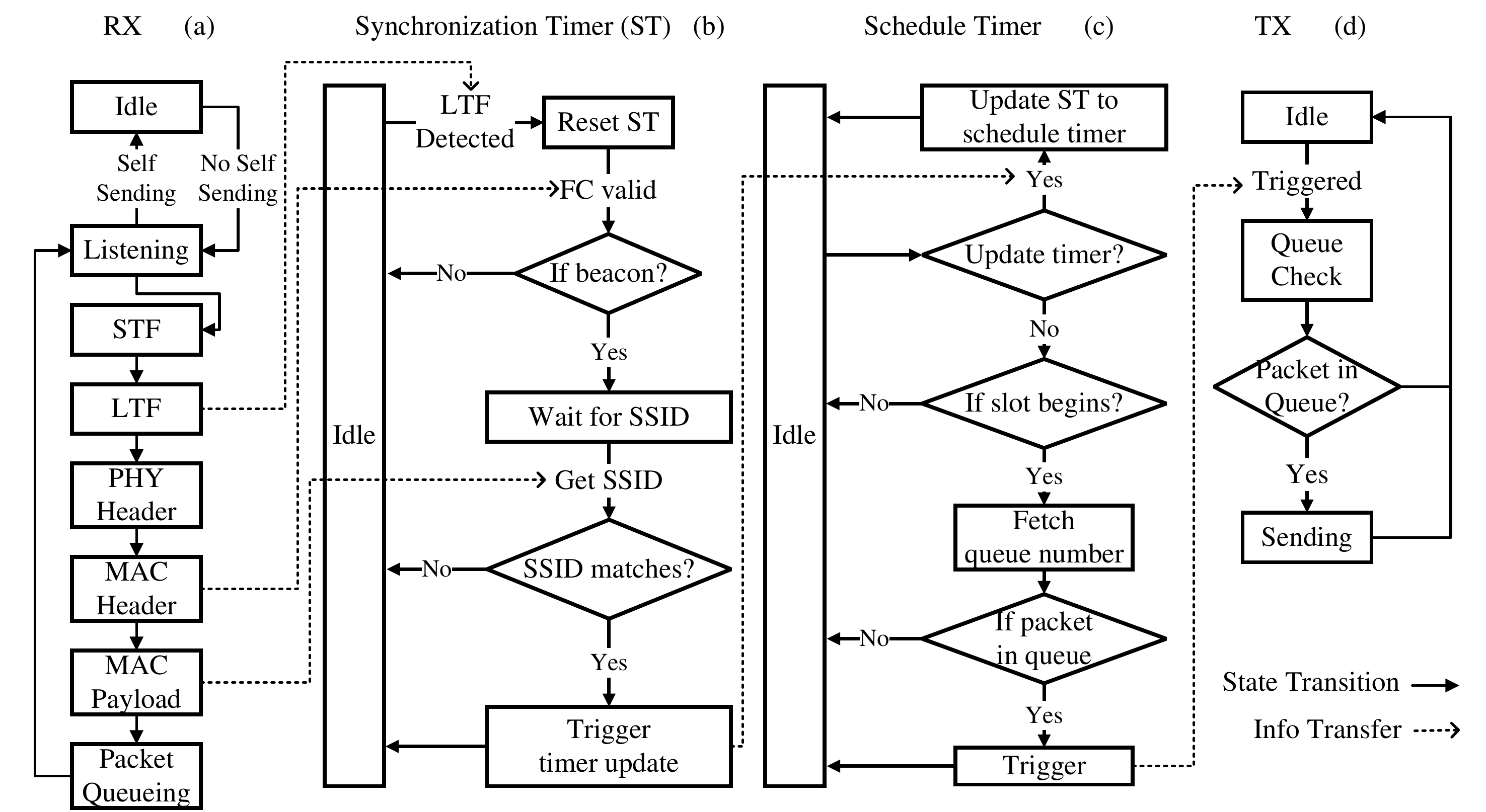}
    \caption{State machines of the schedule timer and the synchronization timer (ST) in the TDMA block of SRT-WiFi.}
    \label{fig:tdmasm}
    \vspace{-0.1in}
\end{figure}

To enable real-time communications in SRT-WiFi and implement the schedule in PL, a TDMA block is added in the XPU module (see Fig.~\ref{fig:arch}). In the TDMA block, a register page is implemented for the TDMA driver.  We assign 16 32-bit registers for the schedule and each slot uses 4 bits to support a schedule of 128 slots. This design can be easily extended to support larger superframes. Based on the schedule information, the TDMA block employs a set of timers to trigger the transmissions. 
At the beginning of a time slot, the TDMA block fetches the link information associated with that slot. Each link has its assigned queue in the TXI module, which will be triggered to send a frame if there is any in the queue. The transmission is triggered by one PL clock pulse. The frame is then read by the OFDM TX module and the bit stream is modulated to digital signal stream and processed by DUC. The final digital signal is passed to the DAC interface and emitted from the antenna of the radio terminal.

\subsection{TDMA Time Synchronization Design}

Another key function of SRT-WiFi is to achieve precise time synchronization among the devices in the network. In our design, we have multiple clusters in the same SRT-WiFi network. Each cluster consists of an AP and multiple stations, and the clusters may share the same channel. For those clusters operating on the same channel, the devices need to be well synchronized to avoid potential collision. The synchronization mechanism of existing work using COTS hardware~\cite{leng2019network} is to connect and synchronize the APs through an Ethernet backbone network using the IEEE 1588 protocol~\cite{eidson2006measurement}. The stations are then further synchronized with the APs using the beacon packets. For the synchronization among the APs, the OS of each AP first updates its system timer with IEEE 1588 and then updates the timer in its wireless adaptor, which is used to send packets at run time. For the synchronization on the station side, they listen to the beacon packets and update the system timer. Since the time synchronization on both APs and stations are done by non-real-time OS, it may cause an average time drift between the devices as high as 20 \SI{}{\micro\second}~\cite{wei2013rt}.

To address this problem, in SRT-WiFi, we propose a new synchronization method based on SDR which is performed at the physical layer (PHY). It is worth noting that this method is only suitable for the devices operating on the same channel. For two APs operating on different channels to synchronize, IEEE 1588 will still be employed. For the APs operating on the same channel, we first designate a master AP (MAP) and let the other APs be the slave APs (SAPs). We assume that all SAPs can hear from the MAP, which provides the reference clock. The SAPs synchronize with the MAP, and all the stations synchronize to their corresponding APs.  The key design goal of SRT-WiFi synchronization is to avoid using the timer in non-real-time OS but leverage the timer in hard real-time PL. For this aim, timers are added in the TDMA block with nanosecond precision for synchronization. We call them TDMA timers, and they are set and run in hard real-time. TDMA timers on MAP are set by its OS to unify the time on MAP. TDMA timers on SAPs and stations synchronize with the TMDA timers on MAP using PHY beacon signal and their OS time are synchronized accordingly. 

\begin{table}[h]
  \caption{PDR measurements with different payload sizes and the corresponding slot lengths.}
  \label{tab:slotlen}
  \begin{tabular}{|@{\,}p{70pt}|@{\,}p{15pt}|@{\,}p{15pt}|@{\,}p{15pt}|@{\,}p{15pt}|@{\,}p{15pt}|@{\,}p{15pt}|@{\,}p{15pt}|}
    \hline
    Payload (bytes)&50&100&150&200&300&400&500\\
    \hline
    Slot Length (\SI{}{\micro\second})&110&118&126&130&146&162&174\\
    \hline
    Sampling Rate (Hz)&9090&8474&7936&7692&6849&6172&5747\\
    \hline
    PDR (\%)&99.7&99.6&99.6&99.6&99.3&99.7&99.4\\
    \hline
\end{tabular}
\end{table}

We now introduce the synchronization procedures.

In SRT-WiFi, PHY demodulation is achieved in the OFDM RX module in PL. The demodulated symbols are passed to RXI and XPU. In the TDMA block, a synchronization function is added to utilize the baseband signal demodulation to synchronize with a specific AP. More specifically, two TDMA timers are added, one is called schedule timer and the other is  called synchronization timer (ST). ST is used to track the arrival time of a beacon packet and the schedule timer is to run the schedule. When a new packet arrives and the long training field (LTF) of the PHY signal is detected in the OFDM RX module, ST is reset. Next, the synchronization function waits for the DLL packet header to be received from the OFDM RX module. It checks whether the packet is a beacon packet. If so, it continues to wait for the service set ID (SSID) in the packet payload. Once SSID is read, the synchronization function compares it with the target SSID provided by the TDMA driver through the registers. If they match, ST is updated to the schedule timer; otherwise, the synchronization function waits for the next packet and the schedule timer runs as usual with no update. This timer update procedure is summarized in Fig.~\ref{fig:tdmasm} where Fig.~\ref{fig:tdmasm} (a) shows how a packet is received and passes the information to ST and Fig.~\ref{fig:tdmasm} (b) shows how ST synchronizes accordingly and triggers the update of the schedule timer. Fig.~\ref{fig:tdmasm} (c) shows how the schedule timer changes states to update the time or trigger real-time transmissions as shown in Fig.~\ref{fig:tdmasm} (d). It is worth noting that this synchronization method also works with higher bit rates in IEEE 802.11n/ac/ax standards. With this method, our experiments show that the synchronization time drift of the SRT-WiFi devices can be maintained within 500 ns which is much better than the 20 \SI{}{\micro\second} time draft observed on the COTS hardware. This more precise time synchronization can help reduce the guard time which is to avoid collisions between slots due to the synchronization error and support smaller time slot length which further improves the sampling rates. Table~\ref{tab:slotlen} presents the packet delivery ratio (PDR) test results with varied application layer payload sizes, the corresponding slot lengths and achievable sampling rates. The guard time used in the experiments is set at 10 \SI{}{\micro\second}. From the results, we can observe that with a payload size of 50 bytes, the slot length can be set at 110 \SI{}{\micro\second} and the sampling rate can be as high as 9 kHz. The detailed experimental results can be found in Section~\ref{sec:performance}.

\begin{figure}[h]
    \centering
    \includegraphics[width=1\columnwidth]{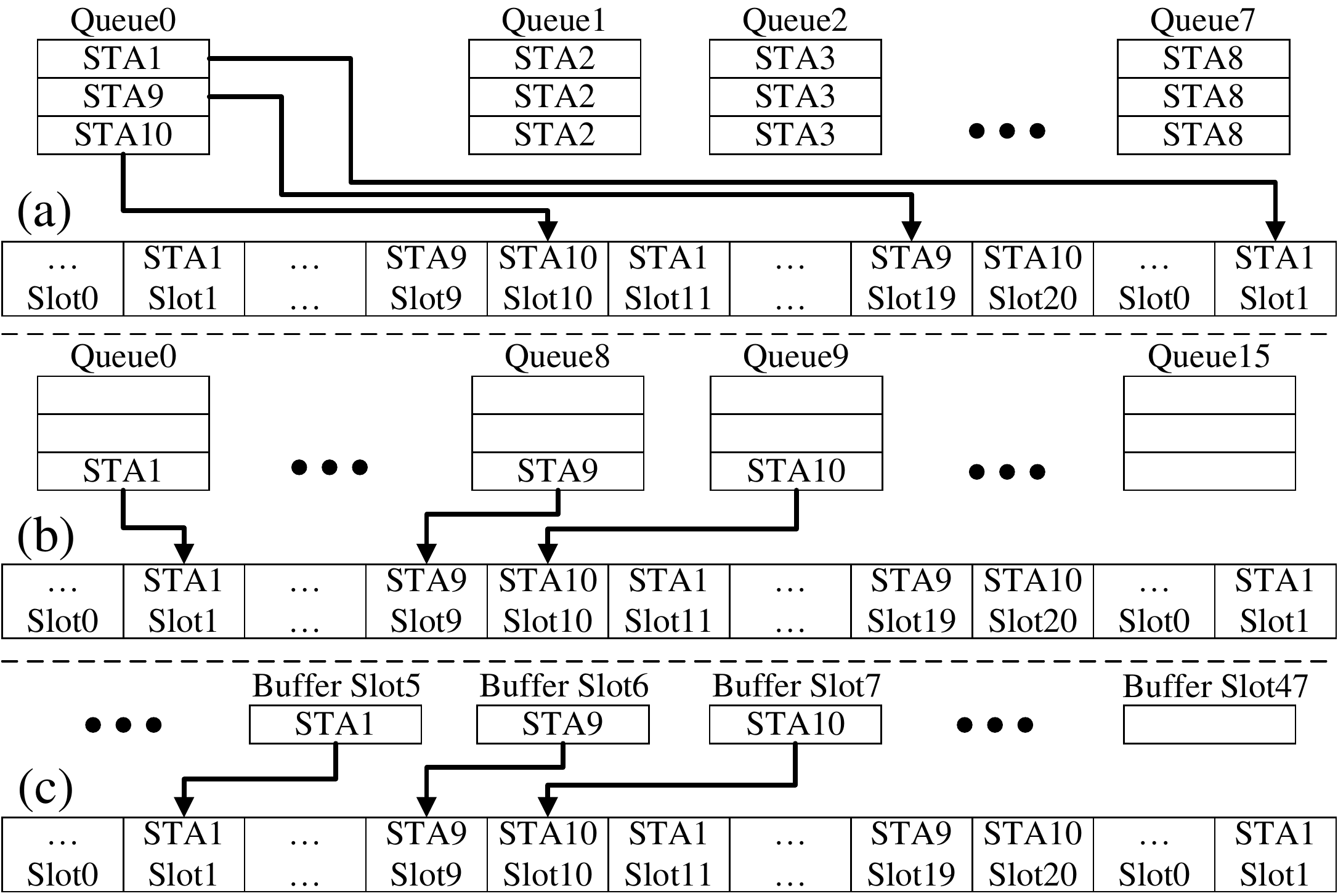}
    \vspace{-0.2in}
    \caption{Queue management issues in real-time networks with shared queues.}
    \label{fig:queue}
    \vspace{-0.2in}
\end{figure}

\subsection{Queue Management}
In SRT-WiFi, before being transmitted, the packets from PS are first pushed into queues. For COTS hardware-based RT-WiFi, the queue implementation is not configurable. For example, AR9285 used in the RT-WiFi implementation~\cite{wei2013rt} uses only 8 queues. To support real-time transmissions in SRT-WiFi, time slots are assigned to individual links to guarantee the desired timing and throughput performance. However, when the number of stations increases beyond the number of queues in the AP, the packets belonging to different links may share a queue, leading to unexpected timing violations. For example, as shown in Fig.~\ref{fig:queue} (a), a device has 10 links while it only has 8 queues so the links of STA9 and STA10 have to share a queue with other links. When two or more packets belonging to different links are pushed into the same queue, they may have to wait until the packet at the queue head being sent, although their assigned time slots in the superframe may come first. This issue happens mainly on the AP side when handling real-time transmissions for multiple stations.

In SRT-WiFi, the packet transmission is triggered by the TDMA block. We assign the queues to different links and the packets belonging to different links are pushed to the corresponding queues as shown in Fig.~\ref{fig:queue} (b). The schedule in the TDMA block stores the information on which queue to be triggered for every slot. This is a feature of SDR-based system since the queues can be defined by software and are not fixed as in the COTS hardware. The number of supported queues can be extended as long as the FPGA has enough resources. 

\begin{table}[h]
    \vspace{-0.1in}
  \caption{Max. and avg. delay (slot number) of packets in assigned and dynamic queue management methods with 16 links.}
  \label{tab:queue}
  \centering
  \begin{tabular}{|c|c|c|c|c|c|}
    \hline
    Number of Queues &8&10&12&14&16\\
    \hline
    Assigned Maximum Delay (slot) &2816&2358&1707&1125&82\\
    \hline
    Assigned Average Delay (slot) &336&236&159&87&16\\
    \hline
    Dynamic Maximum Delay (slot) &591&162&106&104&103\\
    \hline
    Dynamic Average Delay (slot) &271&42&16&16&16\\
    \hline
\end{tabular}
\end{table}

This SDR-based queuing method can eliminate the queuing delay when the number of supported stations is smaller than the number of queues on the AP. However, as the number of stations increases, the queue number cannot be increased infinitely due to the limited resource on the FPGA. To address this issue, we propose a dynamic buffer design in SRT-WiFi as shown in Fig.~\ref{fig:queue} (c) where we use a buffer to replace the previous queues and the buffer is composed of buffer slots and each slot only stores one packet at most. When a packet is passed from the driver, TXI selects an unused buffer slot and pushes the packet into it. At the beginning of each time slot, the TDMA module checks the link information for that time slot. It goes over the buffer to check whether there is a packet belonging to that link. If so, it triggers the transmission of that packet in the corresponding buffer slot. Since a buffer slot only stores one packet, with the same FPGA resources, more buffer slots can be implemented.

Table~\ref{tab:queue} presents the performance comparison between the assigned and dynamic queue management methods with 16 links and the number of queues.  Each link generates a packet periodically and a packet only requires one atomic slot to transmit. If the corresponding queue is available, the packet is pushed into queue. All transmissions follow a randomly generated schedule where the throughput of each link is guaranteed and the length of superframe is fixed. The time between the packet being generated and transmitted is recorded as the packet delay, and packets are not dropped due to the delay.  From the results, we can observe that in the assigned queue management method (where the queues are assigned to links), a few gap between the number of queues and links may cause significantly large max./avg. delay. On the other hand, with the same number of queues, the dynamic queue management method is able to handle more links and keep both max and avg. packet delay small. It however cannot minimize the delay since all the queues are shared. 

\subsection{SNR Measurement}
To support the rate adaptation function in SRT-WiFi, we propose two practical methods to achieve precise SNR measurement in PL. Both methods utilize the short training field (STF) in the preamble of 802.11 PHY signal. The first method computes the cross-correlation~\cite{lee2012digital} of the STF. It is known that STF consists of 10 same short symbols corresponding to 160 samples with 20 MHz sampling rate. So the samples in STF repeat every 16 samples~\cite{80211wg}. After the detection of STF, it is buffered. We use the chips from the \nth{33} to the last one (in total 128 chips) and divide them into two groups each of which has 64 chips. We compute the cross-correlation of the two groups of chips as the $\rho$, and the SNR value (dB) can be computed as follows:
\begin{equation}
\label{equ:2}
\begin{aligned}
\rm{SNR} = 10\log_{10}\left(\frac{\rho}{1-\rho}\right)
\end{aligned}
\end{equation}
where we assume that $\rho<1$. The reason that we use two groups of 64 chips is to exclude the chips at the beginning due to the problems caused by the transient effects of initiating a transmission in the hardware of the sender.

For the second method, after the STF detection, the STF and a piece of background noise before the STF are buffered. The STF signal is added by the background noise. We measure the power of the background noise before the STF and the power of the STF signal which is noise power plus the signal power. Then the SNR (dB) can be computed as:
\begin{equation}
\label{equ:1}
\begin{aligned}
{\rm{SNR}} = 10\log_{10} \left(\frac{P_{\rm{STF}} - P_{\rm{noise}}}{P_{\rm{noise}}}\right)
\end{aligned}
\end{equation}
where $P_{\rm{STF}}$ is the signal power of the STF part and $P_{\rm{noise}}$ is the power of the background noise signal before the STF.  We assume that $P_{\rm{STF}}$ is larger than $P_{\rm{noise}}$.

Both SNR measurement methods are implemented in the OFDM RX module in SRT-WiFi. An SNR value is computed every time when a packet arrives. The computed SNR value is buffered and when the RX interruption in the MAC80211 driver is triggered, it is read through the register. In addition to the SNR value, the driver also needs to know which link the SNR value belongs to since there could be multiple packets being processed during that time. To solve this problem, the SNR value is buffered together with the source address of the packet if applicable (not all the packets have the source address, if not the SNR value is discarded). This information is read in the MAC80211 driver and passed to the TDMA driver and obtained by the device manager on each device. The device manager then sends the SNR information to the central network manager to determine the data rate of each link and the corresponding schedule in the network. The performance of both SNR measurement methods are discussed in Section~\ref{sec:performance}.



\section{SRT-WiFi Processing System (PS) Design}\label{sec:ps}

In this section, we introduce the Processing System (PS) design in SRT-WiFi including the drivers in the kernel and the network managers in the user space. We elaborate how the PS component of SRT-WiFi configures the communication schedules, assigns the queues to individual links and forwards the measured SNR values to CNM for the network management-related decision-making.

\subsection{SRT-WiFi Drivers}
The drivers in SRT-WiFi are the interface between PL and OS (see Fig.~\ref{fig:arch}). They have two main purposes. The first one is to configure parameters in the PL modules to support different working modes and functions; The other purpose is to handle the packet exchange between PL and OS.

We first present the PL configuration and the structure of the driver. As shown in Fig.~\ref{fig:arch}, each module in PL is connected to a corresponding driver in the kernel. The two DMA modules for TX and RX are controlled by the TXI and RXI separately. The DUC and DDC modules are also controlled by TXI and RXI, respectively, to receive parameters from them. The reason to have this structure is that there is a register page in each module in PL. Each register is used by the module to set parameters or read results from the module. For example, TXI uses a register to check if the packet needs an ACK and uses another register to report whether the packet is sent successfully. On the kernel side, each module in PL corresponds to a driver. We call these sub-drivers. They realize the configuration functions by reading and writing the registers in the modules in PL and encapsulate these functions into APIs to be called in the MAC80211 driver. For the TDMA block in XPU, it also has a register page and we divide the registers into three parts. The first part is allocated for the schedule to keep information for the time slots, along with the superframe length and the atomic slot length. The second part is for synchronization purpose since PL needs to know the SSID of the AP that it synchronizes with. The third part is a mode switch. It switches between the CSMA mode and the TDMA mode upon request. To configure these registers, we add a TDMA driver in the kernel. Since the functions of the TDMA mode are not compatible with the MAC80211 subsystem, it is difficult to configure the TDMA block through MAC80211. Instead, we make the TDMA driver a miscellaneous character driver (MISC). It provides the basic reading and writing functions for the user space. In the user space, the network manager calls the APIs of the TDMA driver to configure the TDMA block so that it can modify the schedule, set the parameters and switch the working mode when necessary.

When the MAC80211 sub-system sends a packet, the packet is passed to the MAC80211 driver and handled by the TX operation function (TXO). TXO first checks the priority of the packet and then selects the corresponding queue. Next, TXO determines the TX rate for the packet and checks whether the packet needs RTS/CTS. This information is configured to TXI through the registers. At the end, the packet is fired to the queue in TXI through DMA. In SRT-WiFi, we specify the TX rate and queue for each link. The MAC80211 driver has access to the TDMA driver and TXO fetches the assigned queue and TX rate for the current packet using the destination address of the packet as the key. The queue assignment and rate selection are decided by CNM and stored in a table in the TDMA driver. In our implementation, on the AP side, we use queue 0 for beacon packets and queue 1 for shared links and other broadcast packets. The other queues are assigned to the associated stations. On the station side, it only has the uplink to the AP and a broadcast link so only two queues are needed.

On the receiving side, when a packet is received and queued in RXI, it is forwarded to the kernel through the DMA module. After the packet reaches the kernel, the RX interruption in the MAC80211 driver is called. It checks the information of the packet and forwards it to the MAC80211 sub-system. This indicates that a packet arrives so the SNR value is read from the register in this RX interruption. The SNR value and the corresponding address information are then put into a buffer in the TDMA driver. The device manager finally reads these information and sends them to CNM, which decides and updates the TX rates and schedules for individual links.

\subsection{Network Manager}

In SRT-WiFi, we have three types of network managers forming a network management hierarchy, including the central network manager (CNM), cluster managers (CM) running on the APs and device managers (DM) running on the stations.

For the joining process, when an SRT-WiFi network starts, CNM runs first and waits for CMs to create TCP connections to the CNM and obtain the schedules to be assigned to the links. CMs then start the cluster networks and the slave APs synchronize with the master AP in the same channel and wait for the stations to connect. For the convenience of synchronization and the joining process, all beacon slots and shared slots are fixed during the system operation, and this information is shared with all APs and stations. When a station is powered on, it scans the channels, synchronizes with the designated AP and joins the network. After joining the network, the DM running on the station creates a TCP connection with the CM on the AP to obtain and update the schedule. Before receiving the schedule, the station can only use the shared slots to complete the joining process. Different from the assigned slots, the shared slots are {contention-based}. In a shared slot, each sender first runs a random backoff mechanism as in the CSMA mode and then senses the channel. If the channel is available, the sender then sends the packet. The maximal contention window of the random backoff is configurable. During the system running, DMs and the CMs on each device measures the link qualities and interference. This channel information from each device is obtained by the CNM and it determines and updates the schedules and data rates for the DMs and CMs to adapt the current channel condition so that the links provide stable transmissions.

A unique feature of SRT-WiFi network management is to enable dynamic slot length in the schedule to support run-time rate adaptation. For an individual link, the maximum transmission unit (MTU) is fixed while the data rate may change along with the interference level. With a lower data rate, a packet with the same length requires longer time for transmission which may exceed the boundary of a time slot and cause the collision. In this work, we apply dynamic slot length in the schedule to solve this issue. In the schedule, we define an atomic slot (AS) to be a slot that has the minimum length to support transmitting a packet with a size of MTU at the highest rate. For a packet to be transmitted with a lower rate, it can use multiple consecutive atomic slots in a non-preemptive fashion. Thus by choosing different rates during run time, the packet transmission can take different number of atomic slots. We call this method dynamic slot size assignment. With this mechanism, it is important to compute the schedule for the multi-cluster SRT-WiFi network with rate adaptation and dynamic slot size assignment. In the following section, we will formulate this problem, study its complexity and present a heuristic but effective solution.

\section{Network Management}\label{sec:schedule}

We now formulate the multi-cluster SRT-WiFi network scheduling problem with rate adaptation and prove its NP-hardness. We then present the design detail of a heuristic scheduler to solve this problem. 

\subsection{System Model}

Consider a set $C=\{C_1, C_2, ..., C_m\}$ of clusters in a multi-cluster SRT-WiFi network. Each cluster consists of one SRT-WiFi AP and multiple SRT-WiFi stations forming a star network topology. As the SRT-WiFi network is a time-slotted system, we define an \textit{atomic slot (AS)} as the minimal uninterruptible time unit in the system. For each packet to be transmitted in the SRT-WiFi network, it takes one or multiple  \textit{transmission units}. A transmission unit is configured to be one or multiple atomic slots based on the selected data rates.

Let $\Pi_i = \{\tau_{i,1}, \tau_{i,2}, ..., \tau_{i,n} \}$  be a set of tasks that transmit the packets periodically in cluster $C_i$. Each task $\tau_{i,j}$ is characterized by
\begin{equation}
    \tau_{i,j} = (B_{i,j}, U_{i,j}, D_{i,j}, T_{i,j})
\end{equation}
where $B_{i,j}\in \mathbb{N}$ represents the size of the transmission unit for $\tau_{i,j}$ (in number of atomic slots), $U_{i,j}$ is the number of transmission units required by $\tau_{i,j}$. The deadline and period of $\tau_{i,j}$ are denoted as $D_{i,j}$ and $T_{i,j}$, respectively.

We assume that each task $\tau_{i,j}$ is released in a periodic fashion with a set of instances $\{\mathcal{I}_{i,j,k}\}_{k=1}^{\infty}$. For a transmission unit $l\in[1, U_{i,j}]$ of an instance $\mathcal{I}_{i,j,k}$, let $s_{i,j,k,l}$ and $f_{i,j,k,l}$ represent its \textit{start time} and \textit{finish time}, respectively. Accordingly, let $r_{i,j,k,l}$ and $d_{i,j,k,l}$ be the release time and deadline of the $l^{th}$ transmission unit of $\mathcal{I}_{i,j,k}$. The release time $r_{i,j,k,1}$ of the first transmission unit is the release time of $\mathcal{I}_{i,j,k}$ and the deadline $d_{i,j,k,U_{i,j}}$ of the last transmission unit is the deadline of $\mathcal{I}_{i,j,k}$. In addition, it holds that $r_{i,j,k,p} = f_{i,j,k,p-1}$ with $p\in[2, U_{i,j}]$ and $d_{i,j,k,q} = d_{i,j,k,q+1} - B_{i,j}$ with $q\in[1, U_{i,j}-1]$.

We assume that the sizes of atomic slots for all the clusters are the same, and the number of available channels in the SRT-WiFi network is $H$. We assign each cluster $C_{i}$ to a channel $h_i\in[1, H]$, and introduce the \textit{conflict condition} as follows.

\begin{definition}
For any two instances $\mathcal{I}_{i,j,k}$ and $\mathcal{I}_{i',j',k'}$ with $i = i'$ or $h_i = h_{i'}$ hold, we say that they conflict with each other if the following condition satisfies:
\begin{equation}\label{eq:c1}
    [s_{i,j,k,l}, f_{i,j,k,l}] \cup [s_{i',j',k',l'}, f_{i',j',k',l'}] \neq \emptyset 
\end{equation}
where $l\in[1, U_{i,j}]$, $l'\in[1, U_{i',j'}]$, and it cannot hold that  $i = i'$, $j = j'$, $ k= k'$ and $l= l'$ at the same time. 
\end{definition}

Based on the conflict condition above, we define the feasible condition of scheduling an instance in multi-cluster SRT-WiFi.

\begin{definition} 
For any instance $\mathcal{I}_{i,j,k}$, it is feasibly scheduled in a multi-cluster SRT-WiFi network if it does not conflict with any other instance and the following condition holds:
\begin{equation}\label{eq:c2}
    [s_{i,j,k,l}, f_{i,j,k,l}] \subseteq [r_{i,j,k,l}, d_{i,j,k,l}]  
\end{equation}
where $f_{i,j,k,l} = s_{i,j,k,l} + B_{i,j}$ and $l\in[1, U_{i,j}]$.
\end{definition}

\subsection{Problem Formulation}

The multi-cluster SRT-WiFi network scheduling problem with rate adaptation (MSNS-RA) considers assigning channels to individual clusters and then schedule the transmissions of the packets to eliminate the schedule conflict.

\begin{definition}
Multi-cluster SRT-WiFi Network Scheduling Problem (MSNS-RA): Consider a set of clusters $\{C_i\}_{i=1}^m$ each executing a set of tasks $\{\tau_{i,j}\}_{j=1}^n$ . The MSNS-RA problem is to assign a channel $h_i\in[1, H]$ to each cluster $C_i$ and to find a feasible schedule for all the tasks assigned with rate adaptation on the same channel so that any instance of a task can be feasibly scheduled based on Condition (\ref{eq:c1}) and (\ref{eq:c2}).   
\end{definition}

\begin{theorem}
MSNS-RA is NP-hard in the strong sense.
\end{theorem}
\begin{proof}
Our NP-hard proof uses the 3-Partition problem which is known to be NP-hard~\cite{garey_computers_1990}. An instance of 3-Partition consists of a collection $A = (x_1, x_2, . . . , x_{3n})$ of positive integers such that $\sum x_i = nM$,  $\frac{M}{4}<x_i<\frac{M}{2}$ 
for each $1\leq i\leq 3n$, there exists a partition of $A$ into $A_1$, $A_2$,...,$A_n$ such that $\sum_{x_i \in A_k}x_i = B$ for each $1 \leq k \leq n$~\cite{garey_computers_1990}. 

From any instance of the 3-partition problem, we may construct an equivalent instance of MSNS-RA. Assuming that the total number of available channel in MSNS-RA is 1. Let $\Pi$ be the set of tasks running in all the clusters. For each integer $x_i$, we map a corresponding task in $Pi$ with its size of transmission unit as $x_i$, its number of transmission units as 1, its period and deadline as $M$.  Also, we construct an extra task with its size of transmission unit as $M$, its number of transmission units as 1, its period and deadline as $nM$. The above reduction is a polynomial reduction.

Since the extra task has taken the intervals in $[k \times M, (k+1) \times M]$ where $k\in[0, 2n-2]$ is an even number, 
if there exists a partition of $A$ into $A_1$, $A_2$,...,$A_n$ in the 3-partition problem, we can construct the corresponding solution to schedule each 3 tasks corresponding to the integers in $A_i$ in an unused interval $[(k+1) \times M, (k+2) \times M]$. Also, if we could find a feasible schedule for the MSNS-RA problem, we must schedule every 3 tasks in an unused interval $[(k+1) \times M, (k+2) \times M]$ since $\frac{M}{4}<x_i<\frac{M}{2}$. This shows that there is a feasible schedule if and only if there is a
3-Partition, which proves that the MSNS-RA is NP-hard in the strong sense.
\end{proof}

\subsection{Heuristic Scheduler Design}
To address the MSNS-RA problem in the general case, we propose an effective heuristic scheduler to perform the channel and task assignment. The proposed scheduler design contains a cluster scheduler and a set of task schedulers. The cluster scheduler assigns the channels for individual clusters by balancing the network utilization of channels. Once the channel assignment is completed, the task schedulers are employed to schedule the tasks in each cluster. We now present the details of the two schedulers below.

\subsubsection{Cluster scheduler}

Given $m$ clusters $\{C_i\}_{i=1}^m$, the cluster scheduler first computes and sorts the clusters according to their network utilization in descending order. Specifically, for each cluster $C_i$ with the corresponding task set $\{\tau_{i,j}\}_{j=1}^n$, the network utilization of the cluster is computed as the sum of its task utilization.

To reduce the search space, we introduce a heuristic to assign the clusters to each channel. We define the network utilization of a channel as the sum of the network utilization of all the tasks assigned to this channel. For a cluster $C_i$, with $H$ available channels, we always select the channel with the least network utilization and assign it to $C_i$. For example, we assign $C_1$ to channel 1, $C_2$ to  channel 2, ... , $C_H$ to  channel $H$. For the cluster $C_{H+1}$, we assign it to channel $H$ as the utilization of  channel $H$ is the lowest. For cluster $C_{H+2}$, we check the options of  channel $H-1$ and channel $H$ and select the one with the lowest utilization. 

\subsubsection{Task scheduler}

After assigning the channels for individual clusters, the task scheduler aims to find a feasible schedule for all the tasks assigned to the same channel. Given $n$ tasks $\{\tau_{i}\}_{i=1}^n$ assigned to a channel $h \in [1, H]$ which may be from different clusters with a hyper-period $\mathcal{H}$, we  utilize the release times and deadlines of the transmission units of all the instances of every task from the task set to build the interval set $\mathcal{T}$.  For any interval $I \in \mathcal{T}$ with $I= [s, e]$, $s$ is a release time of a transmission unit and $e$ is the deadline of that transmission unit. Let $\mathcal{D}_{I}$ be the \textit{demand} of the interval $I$, which is defined as the sum of $B_{i,j}$ of any transmission unit of $\mathcal{I}_{i,j,k}$ with its release time and deadline included in $I$. 

Following the EDF (Earliest Deadline First) scheduling policy, we schedule the transmission units based on their deadlines. However, the sizes of transmission units from different tasks might be different and a transmission unit cannot be interrupted during execution. In this non-preemptive case, EDF is known to be non-optimal.  To improve the schedulability, we consider the technique of inserting the idle time.  The key idea is that for each instance popped from the ready queue we utilize the future release patterns of tasks to identify whether or not to insert the idle time to delay its execution. This prevents a non-preemptive transmission unit from being scheduled in an interval such that its demand plus part of this transmission unit becomes larger than the length of the interval, thus jeopardizing schedulability. To overcome this problem, we employ the following rule to insert the idle time in the constructed schedule. Consider any transmission unit of a task instance $\mathcal{I}_{i,j,k}$ to be scheduled at time $t$. If there exists an interval $I= [s, e]$ satisfying the following two conditions:   
\vspace{0.03in}
\begin{itemize}
    \item Condition 1: $[s, e] \subset [t, d_{i,j,k,l}]$
    \vspace{0.03in}
    \item Condition 2: $t + B_{i, j} > e - \mathcal{D}_{I}$.
\end{itemize}
\vspace{0.03in}
then the release time of the transmission unit is set to $s$. 

Since there may exist multiple intervals that satisfy the above conditions, we change the release time of the transmission unit to be the latest one. In addition, deferring the release time to a later time can change the interval set $\mathcal{T}$. We therefore update the interval set if a release time is updated. 

With the above rule to insert idle time in the EDF schedule, we describe the operation of the task scheduler in Alg.~$\ref{alg1}$. The task scheduler first computes the hyper-period $\mathcal{H}$ of the task set assigned to the channel and initializes the interval set $\mathcal{T}$ and the time $t$ based on the timing parameters of the task instances. It then utilizes a ready queue $Q$ to schedule the tasks based on EDF. Specifically, for each instance $I_{i,j,k}$ popped from the ready queue, we employ the rule of inserting idle time to decide if its release time will be deferred (Line 6-11). In addition, we check if the current transmission unit can be scheduled (Line 12-13). If the release time of the current transmission unit is not modified, i.e., $r_{i,j,k,l} = t$, we schedule it in time $[t, t+B_{i,j}]$. Otherwise, we push it back to the ready queue and update the interval set $\mathcal{T}$ (Line 19-20). Let $N=\sum_{i=1}^{n} \mathcal{H}/T_i$ be the total amount of instances of all the tasks  where $T_i$ is the period. As computing the interval set takes $O(N^2)$ time,  the total time complexity of the task scheduler is $O(N^3)$.

\begin{algorithm}[h]
 \DontPrintSemicolon
    \SetKwInOut{Input}{Input}
    \SetKwInOut{Output}{Output}
    \Input{A task set $\Pi=\{\tau_i\}_{i=1}^n$}
    \Output{A schedule $\mathcal{S}$ or reports failure}
    \vspace{0.05in}
    
    Compute the hyper-period $\mathcal{H}$ of the tasks\;
    Initialize a ready queue $Q = \emptyset$, the interval set $\mathcal{T}$ and time $t$\;
    \While{An instance $\mathcal{I}_{x,y,z}$ is released or $Q \neq \emptyset$} {
    
    $Q = Q \cup \{\mathcal{I}_{x,y,z} \}$\;
    Get the earliest instance $\mathcal{I}_{i,j,k}$ in $Q$ at current time $t$ with its $l^{th}$ transmission unit to be scheduled\;
    
    \For {$I \in \mathcal{T}$ with $I= [s, e]$}{
        Compute the demand $\mathcal{D}_{I}$ of the interval $I$\;
        \If {$ I \subset [t, d_{i,j,k,l}] \wedge t + B_{i, j} + \mathcal{D}_{I} > e $} {
            $r_{i, j, k, l} = \max(r_{i, j, k, l}, s)$\;  
        }
    }
    \If {$r_{i, j, k, l} + B_{i, j} > d_{i,j,k,l}$}{
        \KwRet None \tcp{reports failure}
    }
    \uIf {$r_{i, j, k, l} = t$}{
        $[s_{i,j,k, l}, f_{i,j,k, l}] = [t, t + B_{i, j}]$\;
        $t = f_{i,j,k, l}$\; 
    }\Else{
        {insert $\mathcal{I}_{i,j,k}$ to $\Pi$}\;
        update the interval set $\mathcal{T}$\;
    }
    }
    \KwRet $\mathcal{S}$\;   
    \caption{Task Scheduler}
    \label{alg1}
\end{algorithm}

\section{Performance Evaluation}\label{sec:performance}

\begin{figure}[h]
    \centering
    \includegraphics[width=1\columnwidth]{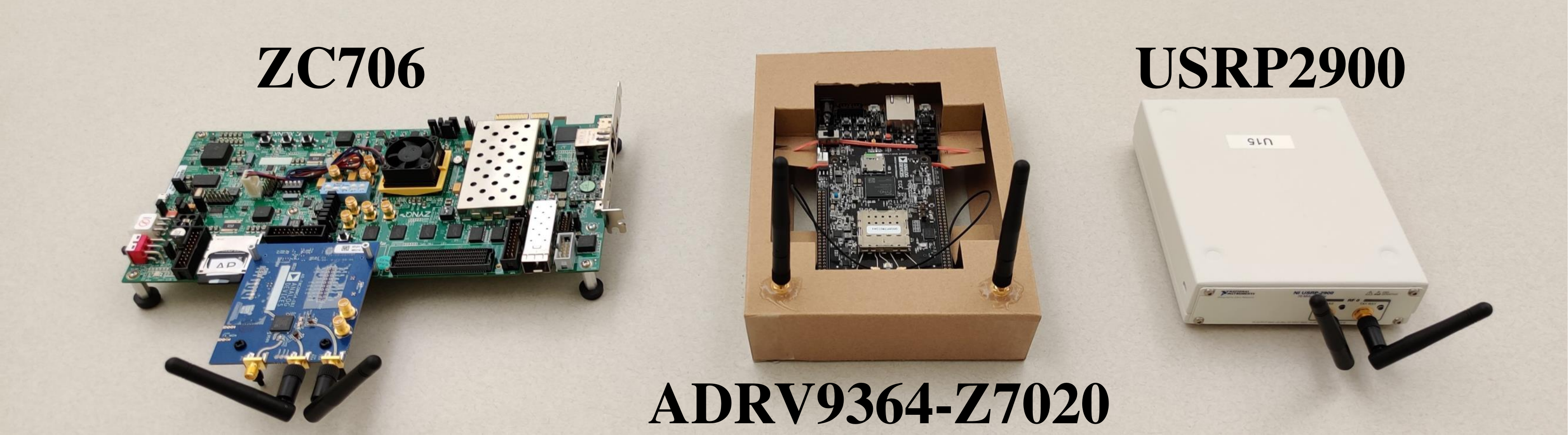}
    \vspace{-0.2in}
    \caption{SDR hardware used in the SRT-WiFi testbeds.}
    \label{fig:device}
    \vspace{-0.1in}
\end{figure}

In this section we report our performance evaluation on the SRT-WiFi design, at both component and system levels.  Fig.~\ref{fig:device} presents the devices used in our SRT-WiFi testbed. We have two hardware platforms. ZC706 consists of Z7045 SoC and AD9364 radio chip and it is used as the hardware for both AP and stations in the SRT-WiFi network. ADRV9364-Z7020 consists of Z7020 SoC and AD9364 radio chip and it is only used for some stations due to its limited FPGA resources. USRP2900 is a traditional SDR device and it is only used for testing purpose in the SRT-WiFi testbed such as signal analysis and interference generation.

\subsection{Synchronization}
We first evaluate the effectiveness of the proposed time synchronization mechanism in SRT-WiFi. To support multi-cluster SRT-WiFi, we let the slave AP (SAP) synchronize with the master AP (MAP) and the stations synchronize with either the MAP or SAP. In the experiments, we first test the beacon interval of the MAP by configuring it to send the beacon packets periodically. We use USRP2900 to capture the beacon signal and use the COTS hardware (AR9285) working in the monitoring mode to sniff the packets. 

\begin{figure}[t]
\centering
  \includegraphics[width=1\columnwidth]{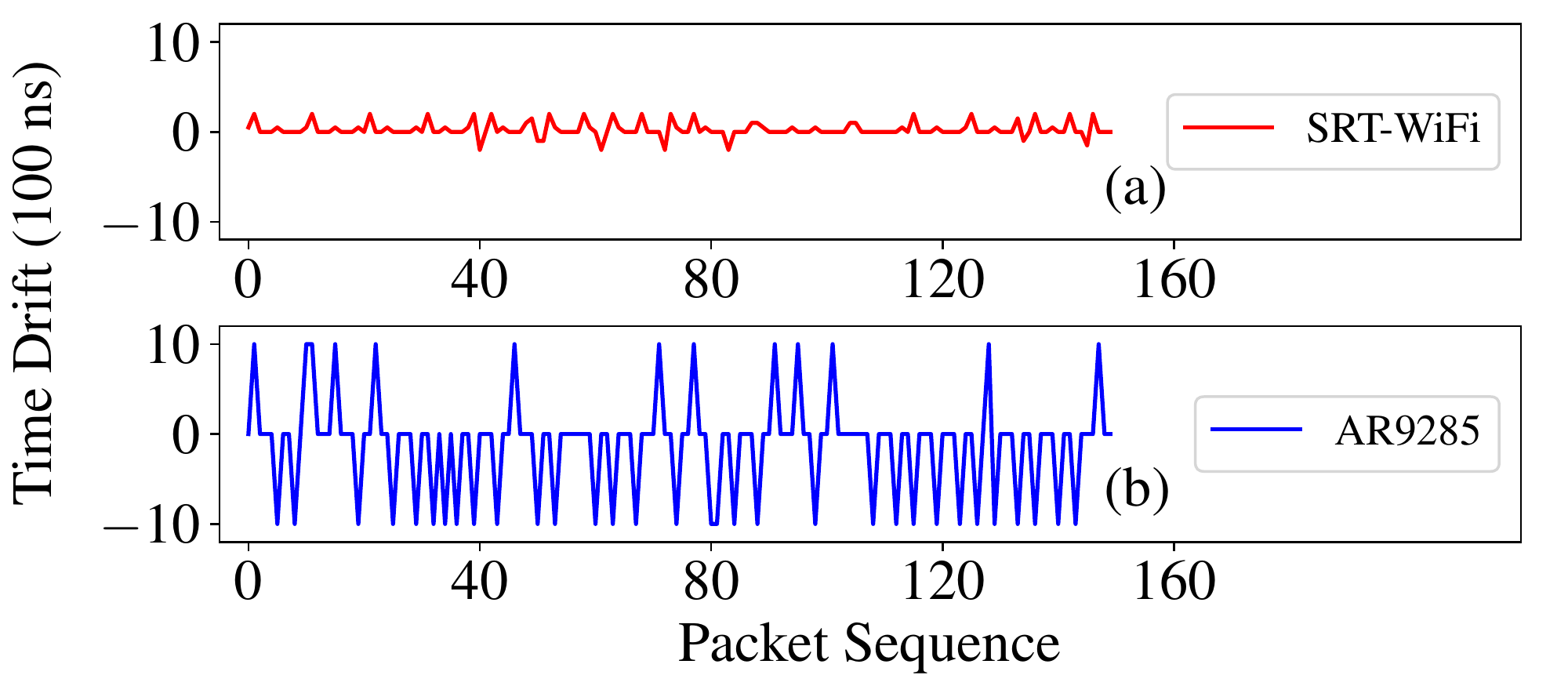}
  \vspace{-0.25in}
\caption{Time drifts in beacon interval by (a) USRP2900 and (b) AR9285.}
\label{fig:beaconinterval}
\vspace{-0.1in}
\end{figure}

In the tests, we set the slot length at 500 \SI{}{\micro\second} and the superframe length at 127 slots so the expected beacon interval is 63.5 ms (one beacon per superframe). Fig.~\ref{fig:beaconinterval} (a) and (b) show the time drift of the beacon interval measured by USRP2900 and AR9285, respectively. The time drift is measured as the error between the inter-arrival time of two consecutive beacons and the expected superframe length (63.5 ms). The SDR result is measured directly from the captured base band signal and the average error, maximal error and the standard deviation are 0.03 \SI{}{\micro\second}, 0.07 \SI{}{\micro\second} and 0.02 \SI{}{\micro\second}, respectively. On the other hand, the result measured from AR9285 has the average error, maximal error and the standard deviation as 0.13 \SI{}{\micro\second}, 0.54 \SI{}{\micro\second} and 1 \SI{}{\micro\second}, respectively. From the comparison, we observed that the timer of SRT-WiFi is much more accurate than that of the COTS hardware. Thus the implementation using COTS hardware needs a larger guard time in the slot design to avoid potential collision.

\begin{figure}[t]
\centering
  \includegraphics[width=1\columnwidth]{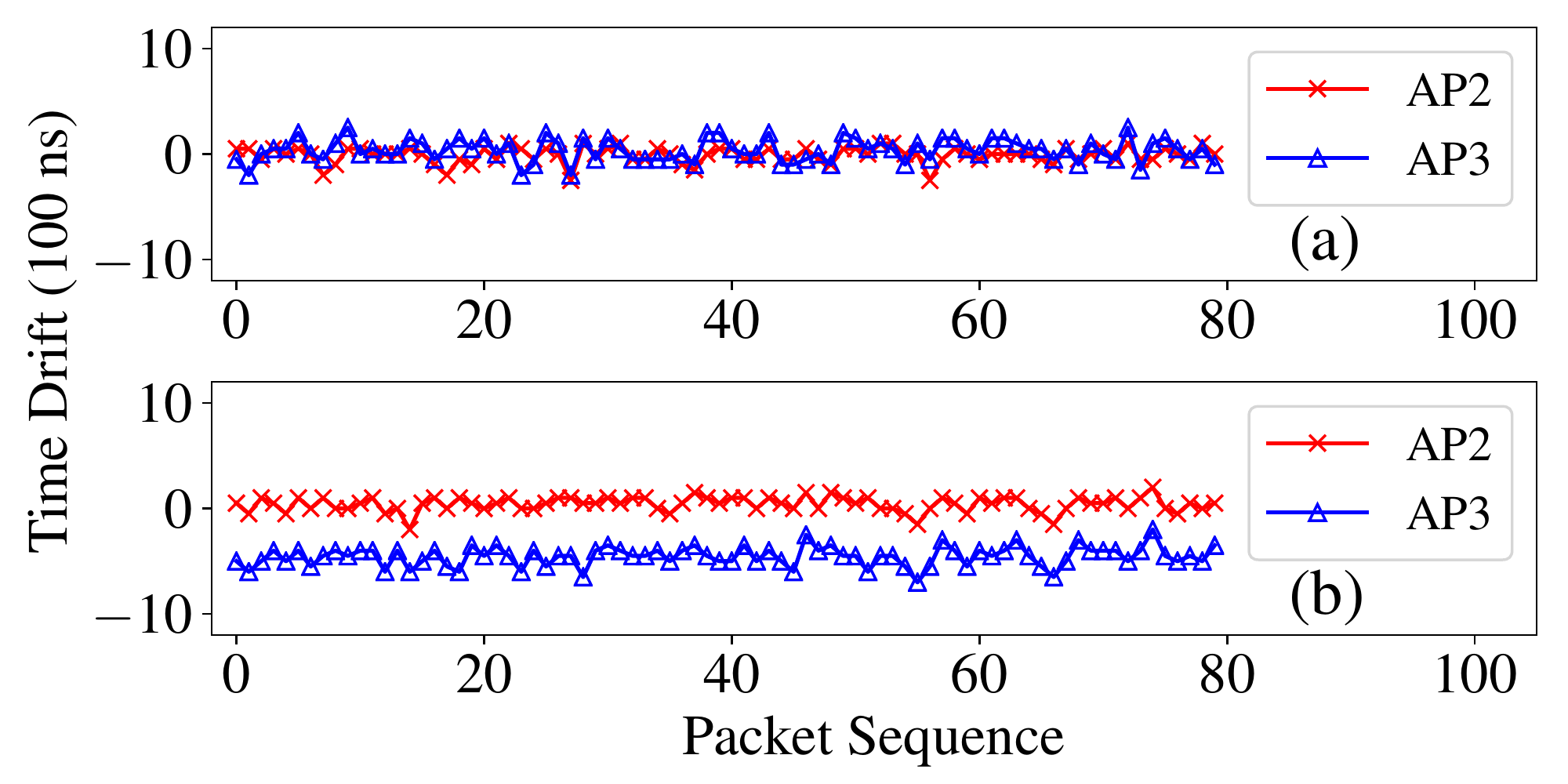}
    \vspace{-0.3in}
    \caption{Synchronization performance of (a) level-1 synchronizations and (b) level-1 and level-2 synchronizations.}
    \label{fig:synctest}
    \vspace{-0.1in}
\end{figure}

Next, we test the synchronization performance of multi-cluster SRT-WiFi networks. In SRT-WiFi, the MAP provides the reference clock. For the SAPs and stations connecting to the MAP, they directly listen to the beacons from the MAP and we call it level-1 synchronization. For the stations connecting to the SAPs, we call it level-2 synchronization. We use three APs for the performance evaluation. To test level-1 synchronization performance, we set AP1 as the MAP sending beacons in slot 0, and AP2 and AP3 as SAPs to synchronize with AP1. We use USRP2900 to measure the beacon sending time of the three APs. Fig.~\ref{fig:synctest} (a) shows the sending time errors of AP2 and AP3. AP2 uses slot 115 to send beacons. The average sending time error from the measurement is 0.01 \SI{}{\micro\second} and the standard deviation is 0.08 \SI{}{\micro\second}. For AP3, it uses slot 117 to send beacons. From the measurement, the average error is 0.03 \SI{}{\micro\second} and the standard deviation is 0.1 \SI{}{\micro\second}. These results show that the accuracy of level-1 synchronization can be well maintained within 1 \SI{}{\micro\second}. Next, we let AP2 synchronize with AP1 while AP3 synchronize with AP2 to test level-2 synchronization performance. We configure AP1 to send beacons in slot 0 and AP2 to synchronize with AP1 and send beacons in slot 2. The measured results in Fig.~\ref{fig:synctest} (b) show that the average error is 0.04 \SI{}{\micro\second}. We further let AP3 synchronize with AP2 and send beacons in slot 119 and the average error is 0.5 \SI{}{\micro\second} and the standard deviation is 0.09 \SI{}{\micro\second}. The results show that although level-2 synchronization is slightly worse than level-1 synchronization, the error can still be within 1 \SI{}{\micro\second} which is significantly better than the COTS hardware. The error comes from not only the SRT-WiFi devices, but the SDR measurement since the sampling rate we use in the experiment is 20 MHz. With a higher sampling rate of SDR, more accurate results are expected.

\subsection{SNR Measurement}

\begin{figure}[h]
\centering
    \vspace{-0.1in}
  \includegraphics[width=1\columnwidth]{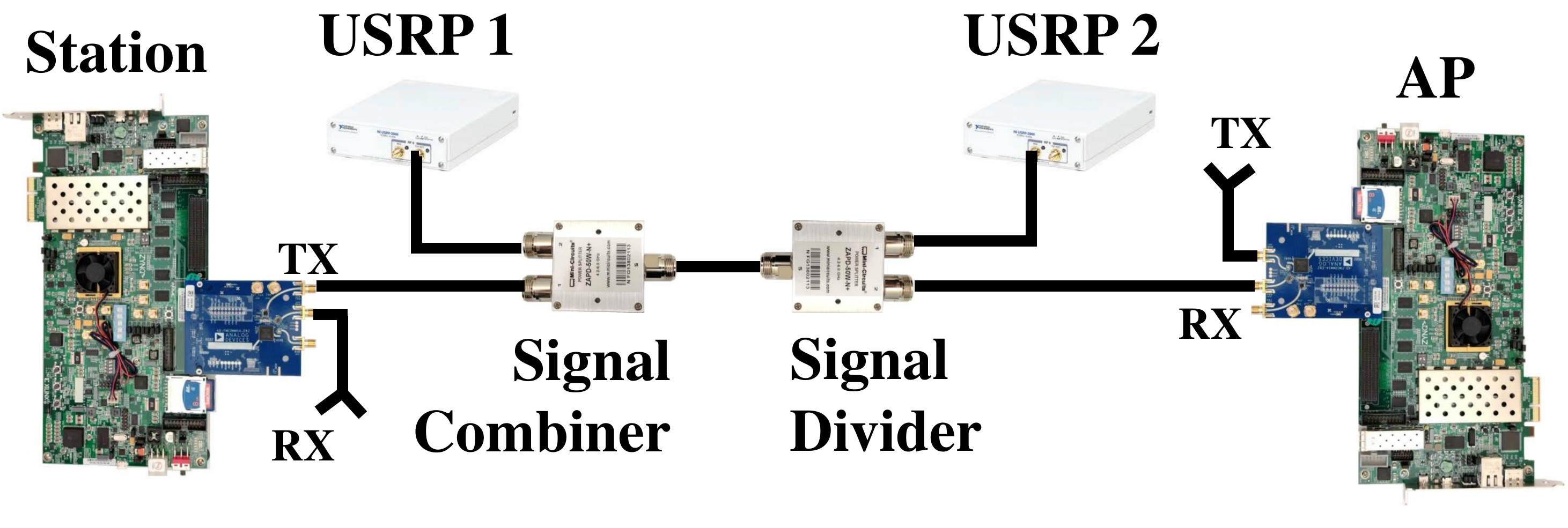}
    \vspace{-0.25in}
    \caption{Setup of the SRT-WiFi testbed for SNR measurement.}
    \label{fig:snrtestbed}
\end{figure}

In the second set of experiments, we test the SNR measurement performance of SRT-WiFi, and the setup of the testbed is shown in Fig.~\ref{fig:snrtestbed}. We use two SRT-WiFi devices, one for AP and one for station. The TX connector of the station connects to the input of a signal combiner. The other input of the signal combiner connects to a USRP device (USRP1). The combined signal is then divided by a signal divider into two ways, one to another USRP (USRP2) and the other connected to the RX of the AP. Both RX of the station and TX of the AP use the antenna. During the experiment, the signal from AP to the station goes on air while the signal from station to AP goes through the cable. We use USRP1 to add controllable noise to the signal from the station to the AP. The AP measures the SNR. At the same time we use USRP2 to record the same signal as the one received at the AP. We then compute the SNR value from USRP2 as the ground truth and compare the results from AP to evaluate the SNR measurement performance.

\begin{figure}[h]
\centering
  \includegraphics[width=1\columnwidth]{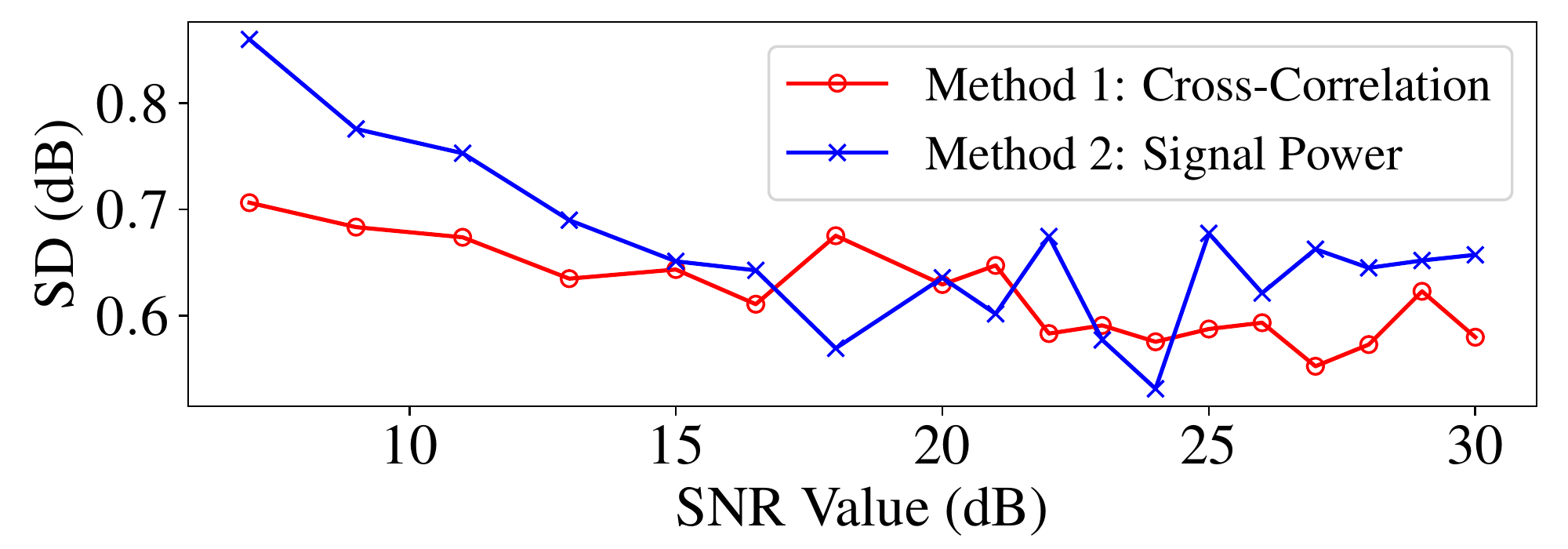}
    \vspace{-0.25in}
    \caption{Standard deviation of two SNR measurement methods measured from the SRT-WiFi testbed.}
    \label{fig:snrmea}
\end{figure}

Fig.~\ref{fig:snrmea} shows the standard deviation measured from both the cross-correlation method and the signal power method. We only test the SNR from 7 dB to 30 dB since when the SNR is lower than 7 dB, the connection between the AP and station is hard to maintain due to the high packet loss rate. The experimental results show that the cross-correlation method outperforms the signal power method in general and thus it is used in all the following experiments.

\begin{figure}[h]
\centering
  \includegraphics[width=1\columnwidth]{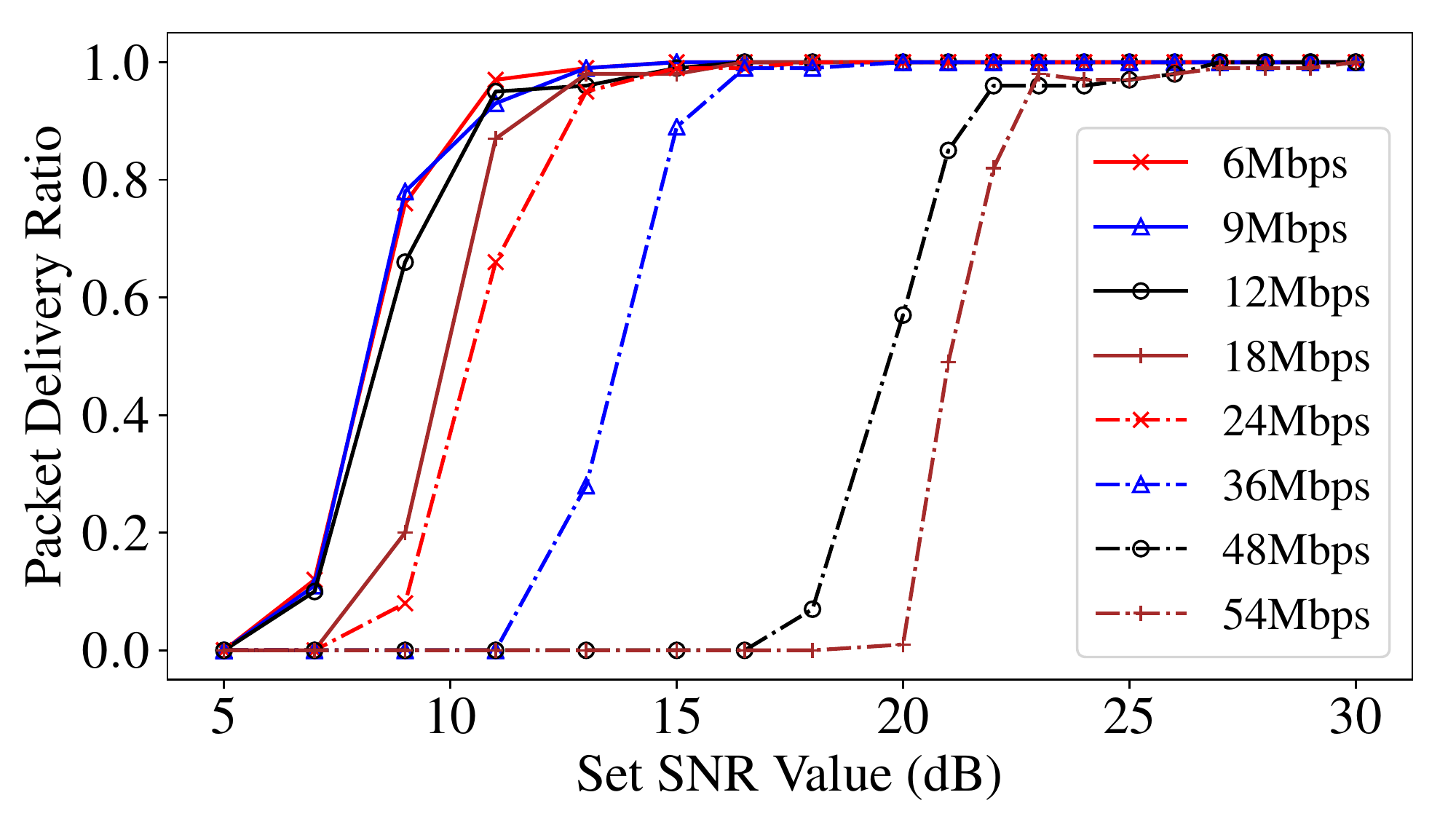}
    \vspace{-0.25in}
    \caption{PDR under different rates against SNR.}
    \label{fig:snrpdr}
    \vspace{-0.1in}
\end{figure}

Fig.~\ref{fig:snrpdr} shows the packet delivery ratios of the SRT-WiFi device under different SNR measured in the testbed. We vary the SNR values by configuring USRP1 to add noise and let the station send UDP packets with a payload size of 500 bytes from the application layer and count the number of received packets on the AP side. This result gives us the reference to perform rate adaptation. 

\begin{table}[h]
    \vspace{-0.1in}
  \caption{TX rates and slot lengths of a 500-byte packet under different SNR values.}
  \label{tab:snrrate}
  \begin{tabular}{|@{\,}p{70pt}|@{ }p{12pt}|@{ }p{12pt}|@{ }p{12pt}|@{ }p{12pt}|@{ }p{12pt}|@{ }p{12pt}|@{ }p{12pt}|@{ }p{12pt}|}
    \hline
    SNR threshold (dB) &25&22&19&17&15&13&10&7\\
    \hline
    Rate (Mbps)&54&48&36&24&18&12&9&6\\
    \hline
    Slot length (\SI{}{\micro\second})&174&186&218&282&342&470&594&846\\
    \hline
    Atomic slot usage&1&2&2&2&2&3&4&5\\
    \hline
\end{tabular}
\end{table}

Table~\ref{tab:snrrate} presents the data rates applied under different channel SNR values and the corresponding slot lengths when transmitting a 500-byte UDP packet. We give the SNR value threshold for each data rate to be used only when the measured SNR value is no smaller than the corresponding threshold. The slot length includes the length of the data packet, the SIFS (16 \SI{}{\micro\second}), the ACK and the guard time (10 \SI{}{\micro\second}).  The settings in Table~\ref{tab:snrrate} are applied in all the following experiments.

\subsection{Throughput and Round Trip Time (RTT)}

In this set of experiments, we evaluate the throughput and round trip time (RTT) of SRT-WiFi, by comparing SRT-WiFi with the regular WiFi using CSMA/CA.

For the throughput comparison, we set up one AP and test different scenarios with one, two and three stations. To find the best system performance, we only use the 54 Mbps data rate. In the experiment, we let the stations keep sending UDP packets with 500-byte payload size to the AP, and the packets received at the AP side are recorded with the corresponding receiving time. For the schedule, we set the superframe to be 127 slots and each slot is 174 \SI{}{\micro\second} (see Table~\ref{tab:snrrate}).


In the throughput test for a single station, we assign 1 slot for beacon, 1 slot for AP and 125 slots for the station to transmit. The expected throughput is 25.52 Mbps and the measured throughput is 25.3 Mbps with a 99.2\% packet delivery ratio (PDR). For the two-station scenario, we assign 1 slot for beacon, 6 slots for the AP and 60 slots for each station. The expected throughput is 12.25 Mbps, and the measured throughput of the first station is 12.18 Mbps with a 99.42\% PDR and that of the second station is 12.251 Mbps with a 99.99\% PDR. It is worth mentioning that the measured throughput of the second station is higher than the expected throughput. This is due to the time drift between the PL timer and the system timer. For the three-station scenario, we still assign 6 slots for the AP and 40 slots for each of the three stations. The expect throughput is 8.17 Mbps, and the measured throughput of the three stations are 8.13 Mbps, 8.12 Mbps and 8.16 Mbps, respectively, with corresponding PDRs as 99.54\%, 99.47\% and 99.94\%, respectively.

We also test three scenarios for the CSMA mode in regular WiFi. For the one-station scenario, it can achieve the throughput of 25.95 Mbps with no packet loss where each packet transmission takes around 174 \SI{}{\micro\second}. This is higher than the theoretical value (25.931 Mbps) which is still due to the timer drift. However, for the two-station scenario, the throughput of the two stations are 3.58 Mbps and 3.62 Mbps, respectively. This is tested with 1 ms UDP packet sending interval. For shorter sending intervals, the CSMA mode is not able to handle it with a large packet loss rate which significantly decreases the throughput. For the three-station scenario, the throughput are 1.88 Mbps,  1.88 Mbps and 1.91 Mbps, respectively.

To evaluate the RTT performance, we also test three scenarios. We let the station send UDP packets to the AP. Once the AP receives a packet, it returns the packet to the sender's address immediately. The station then records the RTT. For the SRT-WiFi network with one station, we set the superframe length at 121 slots and assign 1 slot for beacon, 60 slots for the AP and 60 slots for the station. The AP slots and station slots are placed alternately in the schedule. The average RTT is measured as 752 \SI{}{\micro\second} and the standard deviation (SD) is 105 \SI{}{\micro\second}, and the worst RTT is 1.05 ms. For the two-station scenario, we allocate 30 slots for the AP to each station and another 30 slots for each station to the AP. The average RTT of the two stations are 1.37 ms and 1.41 ms, respectively. The SD values are 203 \SI{}{\micro\second} and 228 \SI{}{\micro\second}, respectively, and the worst RTT is 2.3 ms. For the three-station scenario, the average measured RTT are 1.2 ms, 1.23 ms and 1.22 ms, respectively. The SD values are 410 \SI{}{\micro\second}, 404 \SI{}{\micro\second} and 414 \SI{}{\micro\second}, respectively. The worst RTT is 2.75 ms. It is worth mentioning that the average RTT of the two-station scenario is larger than that of the three-station scenario. This is because after the AP receives the packet, the return packet may miss the first slot from AP to the station due to the processing delay in the OS. For the three-station scenario, the gap between the RX slot for the AP and the TX slot for the stations is larger so the delay is shorter.

Regarding the RTT measurement of the CSMA mode in regular WiFi, the one-station scenario has an average RTT of 508.1 \SI{}{\micro\second}. This is lower than that of SRT-WiFi since in the CSMA mode, the packet can be sent at any time. For SRT-WiFi, packet transmissions strictly follow the schedule. If the sending slot is missed, the packet has to wait for the next one. For the two-station and three-station scenarios, the performance degrades significantly. For the two-station scenario, the RTT measurement only works when the sending interval is larger than 5 ms.  The average RTT is 613 \SI{}{\micro\second} and 706 \SI{}{\micro\second}, respectively. With this setting, for the three-station scenario, the interval is increased to 10 ms, and the RTT values are measured at is 630 \SI{}{\micro\second}, 703 \SI{}{\micro\second} and 685 \SI{}{\micro\second}, respectively.\footnote{It is worth noting that, all the throughput and RTT tests are performed with SDR hardware. For COTS hardware-based regular WiFi system, the performance in CSMA mode may be different~\cite{wei2013rt}.}


\subsection{Rate Adaptation}
\begin{figure}[h]
\centering
    \vspace{-0.1in}
  \includegraphics[width=1\columnwidth]{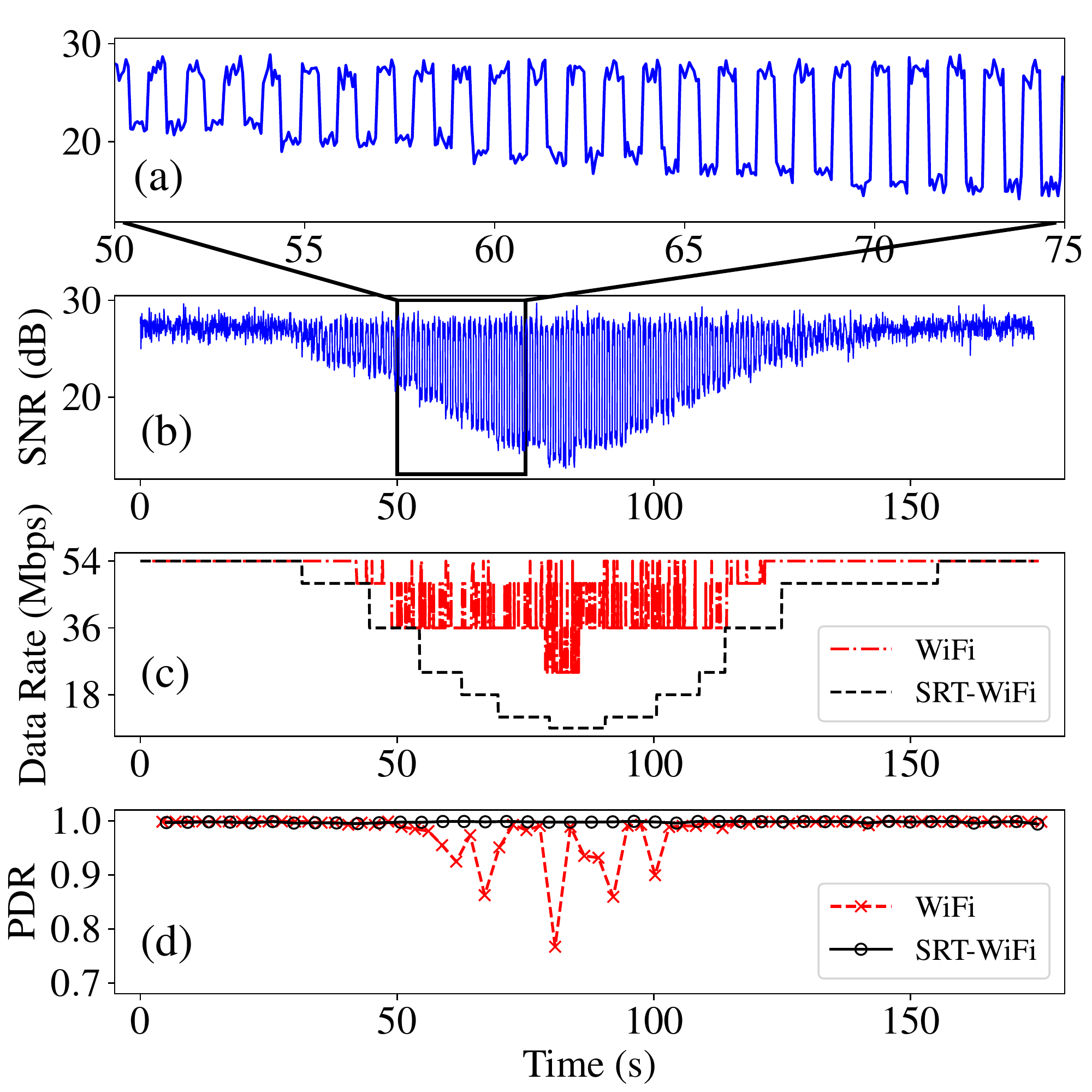}
    \vspace{-0.25in}
    \caption{Data rate and throughput comparison between SRT-WiFi and WiFi in the presence of interference.}
    \label{fig:snrratethrou}
\end{figure}

In this subsection, we demonstrate the effectiveness of the rate adaptation function in SRT-WiFi. In the experiment, we add interference to the channel and measure the data rates and PDR for both SRT-WiFi and regular WiFi networks. In SRT-WiFi, the reception SNR is measured at each device and sent to the central network manager (CNM). CNM then decides the data rate and constructs the schedule for the devices. In this experiment, we set up one AP and one station. We add the interference at the AP side and let the station to send UDP packets to the AP and measure the PDR and SNR. The level of interference is not fixed but varies every 0.5 second, meaning that in the first half of each second, the interference rises to a set level while in the next half of that second, the interference shuts down so that we make the interference change fast. Fig.~\ref{fig:snrratethrou} (b) shows the measured SNR of the channel and Fig.~\ref{fig:snrratethrou} (a) zooms in part of the measured SNR to show how the interference varies. The minimum SNR values first decrease from 27 to 12 dB and then gradually increase back. The data rates of both SRT-WiFi and WiFi are shown in Fig.~\ref{fig:snrratethrou} (c). Regular WiFi uses the Minstrel algorithm~\cite{xia2013evaluation} for rate adaptation, which adapts to the interference according to the transmission history. The corresponding PDR is shown in Fig.~\ref{fig:snrratethrou} (d). From the figure it can be clearly observed that when the SNR value is lower than 20 dB, regular WiFi is not able to keep stable transmissions. In SRT-WiFi, we employ a conservative rate adaptation method. The CNM buffers the measured SNR values for a time window and uses the rate according to the lowest SNR value in the buffer. Once a lower SNR is measured, the data rate is reduced immediately. The rate does not go back until all the SNR values in the buffer are higher than the SNR threshold of a higher rate (see Table~\ref{tab:snrrate}). Although this method wastes some resources when the channel condition is good, it provides stable transmissions. The performance of rate adaptation in SRT-WiFi in the presence of interference is shown in Fig.~\ref{fig:snrratethrou} (c) and it is a step shape without fast changes. Fig.~\ref{fig:snrratethrou} (d) shows the PDR of SRT-WiFi during the test. It is always stable because it measures the lowest SNR and applies the corresponding rate to improve the reliability.

\subsection{Schedule Management}

\begin{table}[h]
    \vspace{-0.1in}
  \caption{Comparison of schedulability among the proposed heuristic scheduling algorithm, EDF and Z3-Solver.}
  \label{tab:schedule}
  \centering
  \begin{tabular}{|c|c|c|c|c|c|c|c|}
    \hline
    Utilization &0.3&0.4&0.5&0.6&0.7&0.8&0.9\\
    \hline
    EDF (\%)&45.8&19.5&10.4&5.2&2.6&0.8&0.5\\
    \hline
    HTS (\%)&75.4&48.6&32.5&19.5&9.8&4.2&1.5\\
    \hline
    Z3 (\%)&75.6&49.4&33.9&21.0&11.1&5.0&1.9\\
    \hline
    EDF \& RCS (\%)&20.9&11.6&6.9&2.8&1.1&0.3&0\\
    \hline
    HTS \& RCS (\%)&30.4&17.7&9.6&4.3&2.1&0.6&0.1\\
    \hline
    HTS \& HCS (\%)&52.7&30.1&15.8&8.1&4.5&1.1&0.1\\
    \hline
\end{tabular}
\end{table}

\begin{table}[t]
\caption{Comparison of the computation cost among the proposed heuristic scheduling algorithm, EDF and Z3-Solver on large task sets.}
\label{tab:schedule2}
\centering
\begin{tabular}{|c|c|c|c|}
\hline
Scheduler & Average time cost (s) & Schedulability & Termination ratio \\
\hline
EDF & 0.019 & 0.75\% & 0\%\\
\hline
HTS & 2.609 & 18.37\%  & 0\%\\
\hline
Z3 & 2732.706 & 5.35\%  & 45.9\%\\
\hline
\end{tabular}
\vspace{-0.1in}
\end{table}

We now present our simulation results along with a case study to show the effectiveness of the proposed heuristic scheduling method to solve the MSNS-RA problem. In the simulation studies, we evaluate the performance of the proposed heuristic task scheduler (HTS) and heuristic cluster scheduler (HCS). For HTS, we compare it with EDF scheduler and an efficient satisfiability modulo theories (SMT) solver Z3~\cite{de2008z3}. All the algorithms including Z3 are implemented in Python and computed in a CPU cluster node with Xeon E5-2690 v3 2.6 GHz CPU. The scheduling problem on each channel is formulated as a constraint programming problem, which could be solved by Z3.

We simulate random task sets to evaluate the schedulability (percentage of schedulable task sets among all the generated ones) of the three methods under the single-channel single-cluster scenario. For each task set, we randomly generate around 10 tasks with the total channel utilization varied from 0.3 to 0.9. Each schedulability value is generated with the simulation of 2000 task sets. The schedulability of the three methods with different channel utilization is shown in row 2 to 4 of Table~\ref{tab:schedule}. The results show that HTS is significantly better than EDF while slightly lower than Z3. In the results we keep the infeasible task sets for comparison to show the trend of how the utilization affects the schedulability. We further compare the time costs of the three methods. Here we generate large-scale task sets with 100 to 150 tasks in each task set and use random channel utilization from 0.3 to 0.9. Each schedulability value is still generated with the simulation of 2000 task sets. The results are shown in Table~\ref{tab:schedule2} and it is clear that Z3 costs much more time than HTS. Besides, we set a timeout for Z3 as 5000 second and it reports 45.9\% terminated cases. The above results show that HTS can achieve a good balance between performance and time cost.

For HCS, we compare it with the random cluster scheduler (RCS) which randomly assigns the clusters to the channels under the multi-cluster multi-channel scenario. In each task set, we randomly generate 4 to 8 channels and 2 to 10 clusters with 5 to 15 tasks in each cluster. The average channel utilization is also varied from 0.3 to 0.9. The tasks of clusters are assigned to channels by HCS or RCS and then each channel is scheduled by HTS or EDF. If all the channels are schedulable we count it as a schedulable case and finally compute the schedulability. From Table~\ref{tab:schedule} row 5 to 7, with RCS, HTS keeps the advantage comparing to EDF. With the proposed HTS and HCS working together, the schedulability is further improved compared with using HTS and RCS together.


\begin{figure}[h]
    \centering
    \includegraphics[width=1\columnwidth]{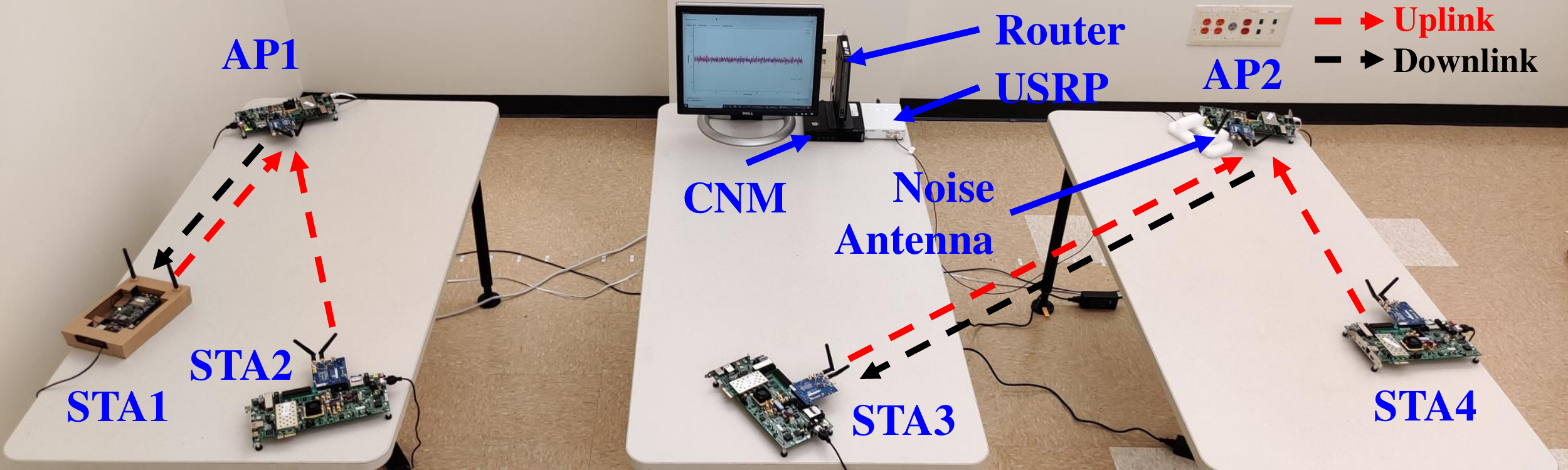}
    \vspace{-0.2in}
    \caption{An overview of the multi-cluster SRT-WiFi testbeds.}
    \label{fig:mctestbed}
    \vspace{-0.1in}
\end{figure}

\begin{figure}[h]
\centering
  \includegraphics[width=1\columnwidth]{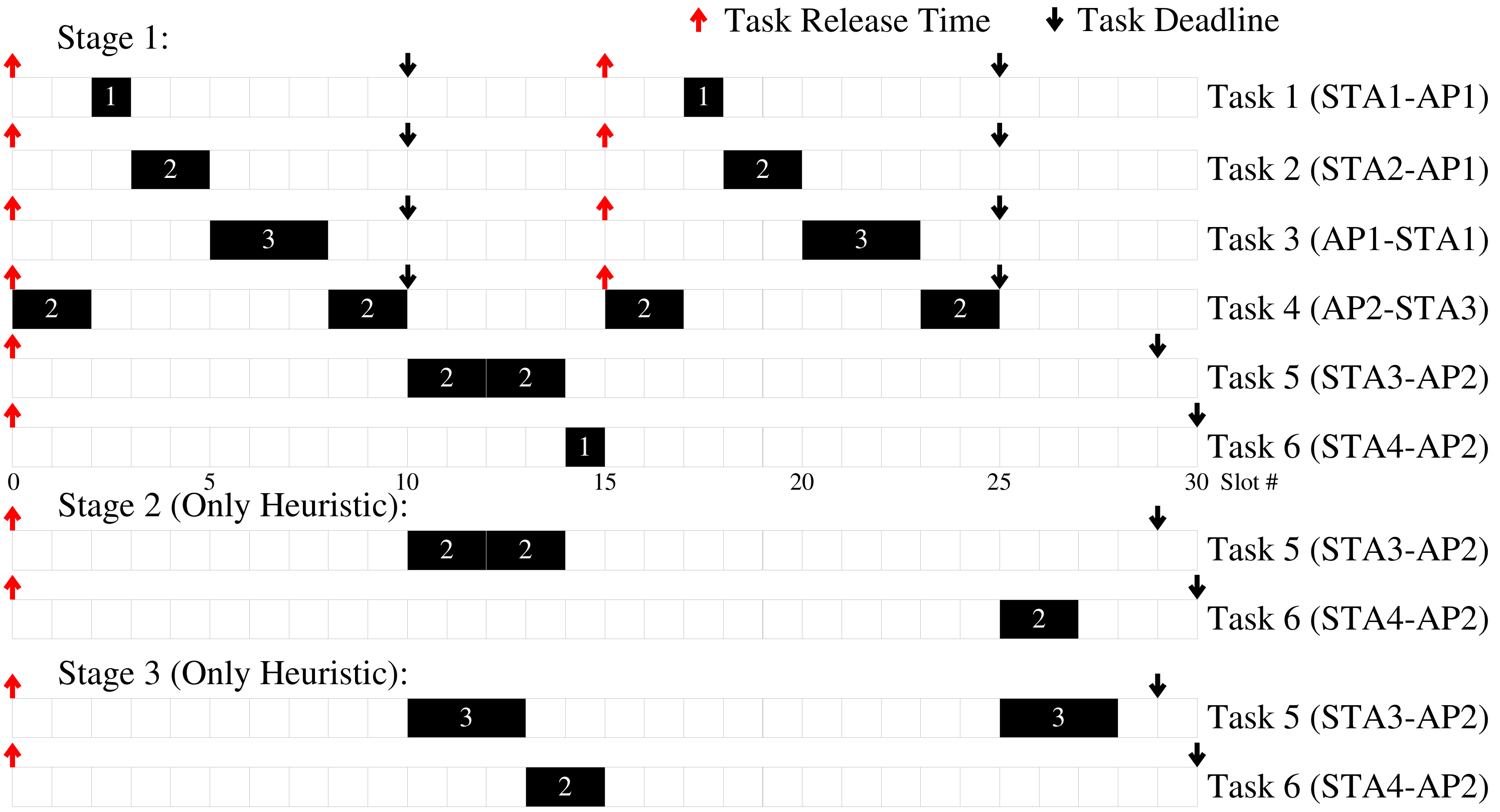}
    \vspace{-0.2in}
    \caption{The task schedule constructed by the MSNS-RA heuristic scheduling algorithm in the case study.}
    \label{fig:taskschedule}
    \vspace{-0.1in}
\end{figure}

\begin{table}[h]
    \vspace{-0.1in}
  \caption{Parameters of tasks used in the case study.}
  \label{tab:taskschedule}
  \centering
  \begin{tabular}{|c|c|c|c|c|c|c|}
    \hline
    Task Number & 1 & 2 & 3 & 4 & 5 & 6 \\
    \hline
    Period (AS) & 15 & 15 & 15 & 15 & 30 & 30\\
    \hline
    Deadline (AS) & 10 & 10 & 10 & 10 & 29 & 30 \\
    \hline
    Transmission Unit Number & 1 & 1 & 1 & 2 & 2 & 1 \\
    \hline
\end{tabular}
\end{table}

In addition to the simulation studies, we also implement HTS on a multi-cluster SRT-WiFi testbed and perform a case study to validate the design correctness. This multi-cluster SRT-WiFi network is configured with two APs (AP1 and AP2 for Cluster1 and Cluster2, respectively) in one channel and each AP is connected with two stations (STA1 and STA2 in Cluster1, STA3 and STA4 in Cluster2). Fig.~\ref{fig:mctestbed} gives an overview of the testbed including the Central Network Manager (CNM), APs and stations with a total number of 6 links. CNM and APs are connected to a router to form a backbone network. A USRP device is further used to generate interference through the noise antenna which is placed next to AP2. As shown in Fig.~\ref{fig:taskschedule}, we assign a task to each link with a specified period, release time and deadline and the task specifications are summarized in Table~\ref{tab:taskschedule}.  Each instance of the tasks sends a UDP packet with a length of 500 bytes. With a fixed packet length, the transmission time of the packet depends on the data rate. In the experiment, we set the atomic slot (AS) length as 174 \SI{}{\micro\second} which is for transmitting a packet at 54 Mbps. The slot lengths of other rates are shown in Table~\ref{tab:snrrate}.

\begin{table}[t]
  \caption{Comparison of SNR, PDR and data rates in the case study.}
  \label{tab:casestudy}
  \begin{tabular}{|@{}p{40pt}|@{}p{16pt}|@{}p{16pt}|@{}p{16pt}|@{}p{16pt}|@{}p{16pt}|@{}p{16pt}|@{}p{16pt}|@{}p{16pt}|@{}p{16pt}|}
    \hline
    Scheduler  & \multicolumn{3}{c|}{SNR (dB)} & \multicolumn{3}{c|}{PDR (\%)} & \multicolumn{3}{c|}{Rate (Mbps)}\\
    \cline{2-10} \& Stage & AP2 & STA3 & STA4 & AP2 & STA3 & STA4 & AP2 & STA3 & STA4\\
    \hline
    Heu.@S1 & 20.8 & 21.1 & 26.2 & \multirow{3}{*}{98.98} & \multirow{3}{*}{98.94} & \multirow{3}{*}{97.24} & 36 & 36 & 54\\
    \cline{1-4}\cline{8-10}
    Heu.@S2 & 20.6 & 15.9 & 19.0 & & & & 36 & 18 & 24\\
    \cline{1-4}\cline{8-10}
    Heu.@S3 & 20.5 & 14.2 & 16.8 & & & & 36 & 12 & 18\\
    \hline
    EDF@S1 & 20.8 & 21.1 & 26.2 & \multirow{3}{*}{98.15} & 98.26 & 98.20 & 36 & 36 & 54\\
    \cline{1-4}\cline{6-10}
    EDF@S2 & 20.6 & 15.9 & 19.0 & & 98.42 & 24.5 & 36 & 18 & 54\\
    \cline{1-4}\cline{6-10}
    EDF@S3 & 20.5 & 14.2 & 16.8 & & 92.31 & 0 & 36 & 18 & 54\\
    \hline
\end{tabular}
\end{table}

In the experiments, STA1 and STA2 in Cluster1 transmit to AP1 in the uplink and AP1 transmits to STA1 in the downlink. In Cluster2, AP2 synchronizes with AP1 and transmits to STA3 in the downlink. At the same time, STA3 and STA4 transmit to AP2 in the uplink. During the experiments, we add interference with three levels as three experiment stages so that links need to adapt to proper rates to achieve good PDR. We first apply the proposed heuristic algorithm and Stage 1 in Fig.~\ref{fig:taskschedule} shows the constructed schedule. We then increase the noise level to Stage 2 and Stage 3. The measured SNR, PDR and applied data rates of links in Cluster2 (sender name is used to mark a link) are summarized in Table~\ref{tab:casestudy} (a PDR of multiple rows is the average PDR). The link quality of STA3 and STA4 drop significantly in each stage because the noise antenna is placed close to AP2. The SNR of STA3 and STA4 drop to 14.2 dB and 16.8 dB, respectively, therefore the rates of STA3 and STA4 drop to 12 Mbps and 18 Mbps, respectively, to address the interference.  On the other hand, STA1, STA2 and AP1 are barely affected by interference and their data rates are 54 Mbps, 36 Mbps and 12 Mbps, respectively. Their average PDR are 98.59\%, 98.53\% and 98.24\%, respectively.

We then evaluate the performance of EDF under the same experiment settings. EDF is only able to generate a feasible schedule in Stage 1 and devices cannot require more atomic slots when the interference level increases.  With this constraint, in Stage 2 and Stage 3, the data rate of STA3 can only drop to 18 Mbps while the rate of STA4 keeps the same. This causes the PDR of STA4 to drop significantly in Stage 2 and 3 and its connection breaks in Stage 3. In Cluster1, the average PDR of STA1, STA2 and AP1 are 99.13\%, 99.22\% and 98.22\%, respectively, and the rates keep unchanged. These results confirm that our proposed heuristic scheduling algorithm can generate feasible schedules and keep higher PDR in more noisy scenarios compared to EDF. And for the schedule updating details, please refer to Section~\ref{sec:ps}.

\section{Related Works}\label{sec:relatedwork}

There are several surveys on high-speed real-time wireless networking~\cite{tramarin2019real, cheng2019adopting}.~\cite{tramarin2019real} gives a survey on existing high-speed real-time wireless networking solutions and rate adaptation methods.~\cite{cheng2019adopting} gives a comprehensive survey and classification of the recent deterministic enhancement approaches in IEEE 802.11 networks including both the widely used real-time mechanisms like IEEE 802.11e EDCA and HCCA and some on-the-way standards like IEEE 802.11ax/aa/ah.

For the existing works on high-speed real-time wireless networking implementation, most of them are soft real-time based on the IEEE 802.11e~\cite{trsek2006simulation, cena2010performance, moraes2007vtp, patti2015schedwifi, peon2016medium, tian2016deadline, costa2019handling}. IEEE 802.11e introduces the hybrid coordination function (HCF) controlled channel access (HCCA) which enables the polling method and the enhanced distributed channel access (EDCA) which enables priorities in the transmissions and uses the highest priority for the real-time transmissions to guarantee their access to the channel.~\cite{seno2012tuning, lam2015fast, lin2015polling, aijaz2020high, zheng2009industrial} study the real-time performance based on the polling method using HCCA.~\cite{lin2015polling} proposes a frequency domain polling method and evaluates the real-time performance through simulations.~\cite{trsek2006simulation, cena2010performance} evaluates the performance of HCCA and EDCA in real-time transmission with simulations.~\cite{moraes2007vtp} claims that the point coordination function (PCF) is not widely implemented on the OTS hardware and it uses the virtual token which is based on the  802.11e EDCA where all the real-time stations are assigned the highest priority in the contention. The real-time stations hold the token and use the channel in turns.~\cite{patti2015schedwifi} proposes a {TDMA-based} MAC protocol based on the IEEE 802.11e EDCA where all the stations follow the schedule and the station in the corresponding slot sends the packets using the highest priority to avoid contention with other stations and the collision.~\cite{peon2016medium} presents three different MAC layer protocols for the real-time wireless communication based on IEEE 802.11e where the first is a scheduled round-robin method and the other two are contention-based and the time critical packets have the highest priority based on EDCA.~\cite{tian2016deadline} presents a deadline-constrained MAC protocol based on IEEE 802.11e by applying a contention sensitive backoff mechanism. ~\cite{costa2019handling} proposes a real-time WiFi architecture which replaces the polling method with a schedule and the real-time traffic is achieved by using the EDCA with the highest priority. Except the IEEE 802.11e, \cite{tramarin2015ieee, tramarin2015use} analyzes the real-time performance of IEEE 802.11n and discusses the advantages of applying IEEE 802.11n on real-time industrial communication.

Some existing works develop their own real-time protocol and systems. \cite{liang2019wia} presents the wireless networks for industrial automation-factory automation (WIA-FA) which is a high-speed real-time wireless solution for industrial automation and factory automation. However, it only supports the ISM 2.4GHz bandwidth and the up link is contention-based which means that the delay from station to the AP is not guaranteed. \cite{wei2013rt, leng2014improving, wei2018schedule, leng2019network} develop a real-time high-speed wireless communication protocol called RT-WiFi. RT-WiFi is a TDMA data link layer protocol based on IEEE 802.11 physical layer to provide deterministic timing guarantee on packet delivery and high sampling rate. It is based on the COTS hardware AR9285 and the upgraded version AR9280.~\cite{romanov2020precise} achieve the time synchronization between devices based on IEEE 802.11 COTS hardware which extend the time sensitive networking (TSN) to wireless networking.~\cite{zhang2021just} also implements a real-time mechanism based on Openwifi, but only focusing on the round-trip time of a two-way request-response communication manner with only a fixed low bit rate. No dynamic rate adaptation or scheduling are considered.

Some works study the rate adaptation for the IEEE 802.11. \cite{holland2001rate} proposes a method for rate adaptation for IEEE 802.11 protocol by using the RTS/CTS before the real transmission where the receiver uses the CTS to send the channel state and rate selection back to the sender and the sender will use the selected rate but the SNR measurement method is not discussed. The performance is evaluated by simulation. \cite{tramarin2016dynamic} proposes a rate selection method for industrial networks based on the knowledge of the SNR of the received packets. In its implementation, the chipset AR9287 has the ability to report RSSI and hence the SNR is able to be computed. \cite{nugroho2013dynamic} proposes a dynamic rate adaptation using the SNR as the threshold to trigger the rate changing and the SNR is measured by the amplitudes of the background noise and the signal. The SNR of the receiver is sent back to transmitter by ACK. The performance is evaluated by simulation.

Except for the solutions based on the IEEE 802.11, there are also some works study the cellular network for industrial real-time applications. \cite{holfeld2016wireless} presents the opportunities that the 5G and LTE can be applied for industrial wireless communication. \cite{ashraf2016ultra} studies the use cases of the 5G in industrial applications through simulations. \cite{aijaz2020private} provides a technical overview of private 5G networks. It introduces the concept and functional architecture of private 5G and the industrial use cases.

Except for the works above focusing on the scheduling part, \cite{suresh2012towards, schulz2014programmatic} build a software-defined WiFi network architecture to provide more flexible network management. The association management is centralized with distributed APs so that it reduces the time cost of station roaming among the APs.

\section{Conclusion and Future Work}\label{sec:conclusion}
This paper presents the design, implementation and performance evaluation of SRT-WiFi, a high-speed real-time wireless system with full stack configurability that is based on software-defined radio (SDR) platform. We discuss the design principles of the programmable logic and processing system of the SRT-WiFi system and show the advantages of SRT-WiFi on high-precision synchronization, queue management and SNR {measurement-based} rate adaptation compared to existing real-time wireless solutions. We further formulate the multi-cluster SRT-WiFi network scheduling problem with rate adaptation (MSNS-RA) and propose an effective heuristic solution to solve it. The performance of the system and the proposed algorithm are thoroughly evaluated in our SRT-WiFi testbed.

For the ongoing and future work, we shall keep improving SRT-WiFi to support evolving features like higher bandwidth, multiple-input and multiple-output (MIMO), beamforming and orthogonal frequency division multiple access (OFDMA) in IEEE 802.11n/ac/ax. Two features under development include 1) higher throughput with 40 MHz bandwidth and single-user MIMO (SU-MIMO) in IEEE 802.11n/ac supporting multiple data streams to be transmitted simultaneously and providing more choices on data rate selection for SRT-WiFi, 2) multi-user multiple-input and multiple-output (MU-MIMO) in IEEE 802.11ac enabling the AP to transmit packets to multiple stations in one time slot which enhances the flexibility of scheduling and further reduces the jitter and latency. Besides the throughput, we also consider more complex network topology like the ad hoc mode. And we will deploy our system in real industrial testbeds to test the performance with interference pattern of real industrial environments. We shall evaluate how these features will affect the resource management in SRT-WiFi to improve the throughput and further reduce the transmission latency in SRT-WiFi networks. 

\section{acknowledgement}\label{sec:ack}
This work was partly supported by the National Science Foundation under Award CNS-2008463 and TI-1919229.

\bibliographystyle{IEEEtran}
\bibliography{references}
\end{document}